\documentclass[aps, twocolumn, groupedaddress, nofootinbib, preprintnumbers,superscriptaddress]{revtex4}
\usepackage{float}
\usepackage[dvips]{graphicx}
\usepackage{color}
\usepackage{amsmath,amssymb,slashed}
\usepackage{tikz}
\usepackage{hyperref}
\usepackage{soul}
\usepackage{tabularx}
\usepackage[normalem]{ulem}
\newcommand{\be}{\begin{equation}}
\newcommand{\ee}{\end{equation}}

\newcommand{\ba}{\begin{eqnarray}}
\newcommand{\ea}{\end{eqnarray}}

\newcommand{\fNL}{f_{\mathrm{NL}}}
\newcommand{\gNL}{g_{\mathrm{NL}}}
\newcommand{\tauNL}{\tau_{\mathrm{NL}}}

\graphicspath{{images/}}

\begin{document}
\title{Post-inflationary Contamination of Local Primordial Non-Gaussianity in Galaxy Power Spectra}
\author{Charuhas Shiveshwarkar}
\affiliation{C. N. Yang Institute for Theoretical Physics and Department of Physics \& Astronomy, Stony Brook University, Stony Brook, NY 11794, USA}
\affiliation{Department of Physics, University of Washington, Seattle, WA 98195, U.S.A.}
\author{Thejs Brinckmann}
\affiliation{Dipartimento di Fisica e Scienze della Terra, Universit\`a degli Studi di Ferrara, via Giuseppe Saragat 1, I-44122 Ferrara, Italy}
\affiliation{Istituto Nazionale di Fisica Nucleare, Sezione di Ferrara, via Giuseppe Saragat 1, I-44122 Ferrara, Italy}
\author{Marilena Loverde}
\affiliation{Department of Physics, University of Washington, Seattle, WA 98195, U.S.A.}
\author{Matthew McQuinn}
\affiliation{University of Washington, Department of Astronomy, 3910 15th Ave NE, Seattle, WA, 98195, USA}
\begin{abstract}
The scale-dependent galaxy bias is an important signal to be extracted in constraining local primordial non-Gaussianity ($\fNL^{\text{local}}$) from observations of large-scale structure. Constraints so obtained rely on the assumption that horizon-scale features in the galaxy power spectrum are exclusively due to primordial physical mechanisms. Yet, post-inflationary effects can induce modulations to the galaxy number density that appear as horizon-scale, scale-dependent galaxy bias. We investigate the effect of two such sources of scale-dependent galaxy bias -- the free-streaming of light relics and fluctuations in the background of ionising radiation -- on precision measurements of local primordial non-Gaussianity 
$\fNL^{\text{local}}$ from galaxy power spectrum measurements. Using the SPHEREx survey as a test case survey reaching $\sigma(\fNL^{\rm local}) \lesssim 1$, we show that ignoring the scale-dependent galaxy bias induced by free-streaming particles can negatively bias the inferred value of $\fNL^{\rm local}$ by $\sim 0.1-0.3\sigma$. Ignoring the effect of ionising radiation fluctuations can negatively bias the inferred value of $\fNL^{\rm local}$ by $  \sim 1\sigma$. The extent of inaccuracies in parameter-inference so incurred depends on the source populations and the ranges of scales used in the analysis, as well as the value of the neutrino mass and the modelling of the impact of ionising radiation. If these sources of scale-dependent galaxy bias are included in the analysis, forecasts for $\fNL^{\rm local}$ are unbiased but degraded. 

\end{abstract}
\maketitle

\section{Introduction}
Cosmic inflation provides a compelling and simple origin of primordial fluctuations in curvature that give rise to the anisotropies in the cosmic microwave background (CMB) and ultimately the distribution of matter and galaxies in the Universe today~\cite{Linde:2007fr,Tsujikawa:2003jp,Guth:1980zm,Turok:2002yq}. In the simplest scenario, the field that drives the exponential expansion of inflation is also responsible for generating primordial perturbations. Yet, in scenarios where additional light fields are present during inflation, primordial perturbations may have a more complex origin. A simple example is the case where primordial perturbations are exclusively sourced by a second light field, different from the inflaton. Example implementations of this are the curvaton scenario \cite{Enqvist:2001zp, Lyth:2001nq, Sasaki:2006kq} and modulated reheating \cite{Dvali:2003em, Dvali:2003ar}; both cases are representative of a variety of scenarios that could arise (for recent reviews, see Refs.~\cite{Meerburg:2019qqi,Achucarro:2022qrl}).

Primordial curvature perturbations inherited from a second light field are generically generated through a local non-linear mapping of the fluctuations seeded during inflation. This non-linearity imprints a specific type of non-Gaussianity on the initial curvature perturbations\footnote{Throughout this paper, we use $f_{\rm NL}$ and $f^{\rm local}_{\rm NL}$ interchangeably.}, 
\begin{eqnarray}
\label{eq:fNLgNLdef}
\zeta(\vec{x}) &=& \xi(\vec{x})+\frac{3}{5}\fNL^{\text{local}}(\xi(\vec{x})^{2}-\langle\xi^{2}\rangle)\\ &&+ \frac{9}{25}\gNL \xi^3(\vec{x}) +\dots \nonumber\ ,
\end{eqnarray}
where above $\zeta(\vec{x})$ is the primordial curvature perturbation and $\xi$ is a Gaussian random field \cite{Salopek:1990jq}. The particular value of the coefficients $\fNL$ and $\gNL$ are model-dependent, but in simple scenarios are $\mathcal{O}(1)$. At present, the most stringent constraints on these parameters come from CMB datasets, with $\fNL = -0.9\pm 5.1$ and $\gNL = (-5.8\pm6.5)\times 10^4$ at $68\%$ confidence from Planck \cite{Planck:2019kim}. As of today, constraints obtained from Large-Scale Structure are significantly less stringent (see, for example,~\cite{Cabass:2022ymb} for recent constraints on $\fNL$ from the twelfth data-release of the Baryon Oscillation Spectroscopic Survey (BOSS)~\cite{Alamata}). In the near term, however, constraints from galaxy surveys are expected to surpass those from the CMB by nearly an order of magnitude \cite{Dore:2014cca,Ferraro:2019uce}.

A key observable in the hunt for primordial non-Gaussianity of the type given in Eq.~(\ref{eq:fNLgNLdef}), is a modulation of the galaxy number density proportional to the primordial curvature perturbation, in addition to the usual modulation by the large-scale matter density field, 
\be
\label{eq:heuristicdeltan}
\frac{\delta n_g}{n_g} \rightarrow \frac{\delta n_g}{n_g} +  b_\zeta \zeta \,,
\ee
where ${\delta n_g}/{n_g} $ is the fluctuation in the galaxy number density and $b_\zeta$ is some coefficient. The equivalence principle forbids a term such as $b_\zeta \zeta$ from being generated dynamically, so this signature can only arise from initial conditions \cite{Creminelli:2011rh}. Since $\zeta$ is related to the matter density fluctuation (in the comoving synchronous gauge)\footnote{Throughout this paper, we work in the comoving synchronous gauge. See footnote \ref{footnote1} (on page \pageref{footnote1}).} as $\zeta\sim \delta \rho_m/k^2$ from Poisson's equation, the imprint of the $b_\zeta \zeta$ term appears as a new term in the galaxy bias that diverges at small wavenumber $k$,
\ba
\label{eq:heuristicbk}
\Delta b \sim \frac{b_\zeta}{k^2}\,.
\ea

Single field inflation does not predict any scale-dependence in the galaxy bias like Eq.~\eqref{eq:heuristicbk} ~\cite{Maldacena:2002vr,dePutter:2015vga}. A detection of non-zero $\fNL$, $\gNL$, or higher order terms in Eq.~(\ref{eq:fNLgNLdef}) would therefore rule out all single-field, slow-roll inflationary models for the origin of structure\footnote{Even the $\gNL$ term predicts a scale-dependent bias that goes as $1/k^2$ on large scales. See, e.g., Ref. \cite{Smith:2011ub}.}. The particular values of these parameters could further be used to infer properties of the additional fields or the process by which the perturbations in additional fields are converted into curvature perturbations. 

While it is indeed the case that post-inflationary astrophysical processes cannot generate a term like Eq.~(\ref{eq:heuristicbk}), there are two known mechanisms that can cause increasing changes to the galaxy bias with decreasing wavenumber, down to values of $k$ approaching the horizon. These are, i) modulations to the galaxy number density caused by fluctuations in ultraviolet (UV) background of ionising radiation \cite{Sanderbeck:2018lwc} and ii) changes to how the CDM modulates the galaxy number density from gravitational interactions with relativistic particles (e.g. neutrinos and photons) \cite{LoVerde:2014pxa, Chiang:2017vuk,Chiang:2018laa, Shiveshwarkar:2020jxr}. Both of these change the galaxy bias on scales related to the propagation distance of relativistic particles, $k \sim aH$, where $a$ is the scale factor and $H$ the Hubble parameter. This violates the conventional wisdom that only inflationary processes could generate correlations that modulate the galaxy number density separately from the matter density at such large scales. 

In Ref.~\cite{Sanderbeck:2018lwc}, it was estimated that UV background fluctuations can modulate the galaxy number density at a level comparable to $\fNL\sim1$, while modelling the UV background contribution to the galaxy power spectrum would degrade constraints on $\fNL$ by $\sim 40\%$. This paper is the first to study the impact of scale-dependent bias from neutrinos and light relics on $\fNL$ forecasts. In this paper, we quantify the importance of scale-dependent bias from neutrinos and the UV background on precision measurements of $\fNL$ from galaxy power spectra, such as targeted by the NASA mission SPHEREx~\cite{Dore:2014cca} and the proposed MegaMapper Survey \cite{Ferraro:2019uce}. SPHEREx is a planned all-sky survey that (among other things) aims to study the imprints of inflationary physics on large-scale structure by probing galaxy redshifts over a large cosmic volume~\cite{Dore:2014cca}. SPHEREx is designed to minimise systematics in the measurement of large-scale galaxy distribution and promises to constrain $\fNL$ tightly ($\sigma_{\fNL}\sim 0.5$). SPHEREx will use both galaxy power spectra and bispectra to achieve their target constraint. In this paper, we restrict our analysis to the galaxy power spectrum coming from the scale-dependent bias, which provides roughly half the constraining power\footnote{Biased tracers, including those used for bispectrum measurements, are generically altered by UV fluctuations and light relics, so constraints on $\fNL$ from the galaxy bispectrum could also be impacted these effects. Yet, much of the constraining power of the galaxy bispectrum for primordial non-Gaussianity is driven by the $\fNL$ contributions to the matter bispectrum that peak in the squeezed limit, rather than the scale-dependent galaxy bias. The squeezed-limit matter bispectrum is only very marginally changed by light relics \cite{Chiang:2017vuk, Shiveshwarkar:2020jxr} and is unchanged by UV background fluctuations. We therefore restrict our attention to the galaxy power spectrum and leave an investigation of the bispectrum of biased tracers to future work.}~\cite{Dore:2014cca}. 

We use the SPHEREx galaxy survey as a test case and build a likelihood that incorporates scale-dependent bias from both UV background fluctuations and neutrinos on the observed galaxy power spectrum. By performing an MCMC analysis, we demonstrate the importance of these terms and their dependence on galaxy populations, redshift, and modelling assumptions. We use the parameter inference package MontePython v3.4~\cite{Audren:2012wb,MontePython3}, and our SPHEREx likelihood is built by generalising the Euclid galaxy power spectrum likelihood~\cite{Sprenger:2018tdb} to multiple galaxy subsamples with the specifications of the SPHEREx all-sky survey~\cite{Dore:2014cca}.

We begin with a review of scale-dependent bias from local-type primordial non-Gaussianity in section~\ref{sec:fNLbias}. In section~\ref{sec:nonPbk}, we review scale-dependent bias from neutrinos and fluctuations in the UV background. In section~\ref{sec:Pk}, we present our model for the galaxy power spectrum in the presence of $\fNL$, neutrinos, and UV background fluctuations. The likelihood used for forecasts is given in section~\ref{sec:likelihood} and our results are presented in section~\ref{sec:results}. We conclude in section~\ref{sec:discussion}. A short appendix~\ref{app:fullspectra} presents the full explicit expression for the galaxy auto and cross-power spectra. 

\begin{table}
\centering
\begin{tabular}{||c|c|c|c||} 
 \hline
 $ \omega_b$ & $\omega_{cdm}$ & $100\theta_s$ & $A_s$ \\ [0.5ex] 
 \hline\hline
 0.02218 & 0.1205 & 1.041126 & $2.032692\times 10^{-9}$\\ [1ex]
 \hline
 \hline
 $n_s$ & $z_{reio}$ & $ M_{\nu}$ & $ \fNL^{local}$\\[0.5ex]
 \hline\hline
 0.9613 & 7.68 & 0.06 eV & 1.0 \\[1ex]
 \hline
\end{tabular}
\caption{Fiducial values for the six standard parameters of the flat-$\Lambda$CDM model, in addition to our fiducial neutrino mass and local primordial non-Gaussianity parameters.}
\label{fiducial_cosmology}
\end{table}

Throughout this paper, we assume a flat $\Lambda$CDM cosmology with massive neutrinos with fiducial parameters as in table \ref{fiducial_cosmology}. Unless otherwise stated, we assume a degenerate hierarchy\footnote{Of course, for small values of $M_\nu$, a completely degenerate mass hierarchy is not consistent with oscillation data, yet this proves to be a good approximation across the range of masses we consider~\cite{Lesgourgues:2006nd}. Additionally, the scale-dependence in the galaxy bias introduced by the free-streaming of neutrinos is mostly sensitive to the total neutrino mass and is not significantly sensitive to the distribution of mass between individual neutrino mass states~\cite{Shiveshwarkar:2020jxr}. See also ~\cite{Archidiacono_2017,CORE:2016npo,Gerbino:2016sgw} for more details. }  of three massive neutrinos with a total neutrino mass of $M_{\nu} = 0.06\ \text{eV}$\footnote{To analyse the impact of neutrino-induced scale-dependent bias, we consider neutrino masses going up to $M_{\nu}=0.3 \text{eV}$, roughly corresponding to conservative cosmological constraints (see, e.g., Ref.~\cite{DiValentino:2019dzu} and~\cite{DiValentino:2021imh}).}. The general strategy we follow is to conduct MCMC runs with our SPHEREx galaxy power spectrum likelihood (plus a mock Planck CMB likelihood, which roughly matches Planck 2018 sensitivity, to break parameter degeneracies~\cite{Brinckmann:2018owf}) with and without the neutrino and ionising radiation effects of different magnitudes and study how this affects the forecast of $\fNL$. This enables us to quantify the extent to which the measurements of $\fNL$ can be biased\footnote{Note that in this paper we use the word `bias' to mean `galaxy bias' or to denote inaccuracies in the measurement of cosmological parameters depending on the context of a given statement.} if these non-primordial effects are not incorporated in modelling the galaxy power spectra. 


\section{scale-dependent Bias due to local Primordial Non-Gaussianity}
\label{sec:fNLbias}

In this section, we present a derivation of the scale-dependent bias caused by local primordial non-Gaussianity. We show that the linear bias of galaxies acquires a scale-dependent correction that diverges at large scales. Throughout this paper, we work in the comoving synchronous gauge because it is only in this gauge that the linear bias relation ($\delta_{g}=b_{g}\times \delta_{c}$) remains valid (at linear order) out to the largest scales~\cite{Jeong:2011as}.\footnote{In other gauges, there are typically additional terms in the linear bias relation that become more important at large scales~\cite{Jeong:2011as}.\label{footnote1}} 

Models of inflation containing additional light fields along with the inflaton typically generate the `local' type of primordial non-Gaussianity as given in Eq.~(\ref{eq:fNLgNLdef}). The canonical example we consider here is of curvature perturbations which are inherited from a single additional light field such as a curvaton~\cite{Smith:2010gx,Tseliakhovich:2010kf}, wherein the primordial curvature perturbation is quadratic in a Gaussian random field as,
\begin{eqnarray}
\label{fNL:def}
\zeta(\vec{x}) &=& \xi(\vec{x})+\frac{3}{5}\fNL(\xi(\vec{x})^{2}-\langle\xi^{2}\rangle)\ ,
\end{eqnarray}
where $\xi$ is a Gaussian random field such that the power spectrum of primordial curvature perturbations is 
\begin{eqnarray}
\label{eq:1.5}
P_{\zeta}(\vec{k}) &=& P_{\xi}(\vec{k}) +\mathcal{O}\left(\fNL^2 P_\xi^2\right)\ .
\end{eqnarray}
In Fourier-space, Eq.~\eqref{fNL:def} implies that the non-Gaussian correction to a Fourier mode is a convolution of the Gaussian curvature \begin{eqnarray}
\label{PNG:fourier}
\zeta(\vec{k}) = \xi(\vec{k}) + \frac{3}{5}\fNL\xi\ast\xi(\vec{k}) - \frac{3}{5}\fNL\langle\xi^2\rangle(2\pi)^{3}\delta^{3}_{D}(\vec{k})    \ .
\end{eqnarray} 
To analyse the effects of local-type primordial non-Gaussianity on the clustering of dark matter haloes, we follow the peak-background-split approach \cite{Bardeen:1985tr}. This approach proceeds with the observation that one can split the density field into a large-scale component, one that only varies significantly over lengths much larger than the radii of typical dark matter haloes, and a small-scale component that eventually undergoes non-linear gravitational collapse to form haloes. The halo bias can then be studied by examining the growth of small-scale matter density perturbations within a `background' of an ambient large-scale matter density. Following in this spirit, one can separate the small and large-scale parts of $\zeta$ and $\xi$ ($\zeta = \zeta_L + \zeta_s $ and $\xi = \xi_{L}+\xi_{s}$) and express the primordial curvature $\zeta$ (up to linear order in all the large-scale fields $\zeta_L$ and $\zeta_s$) as 
\begin{eqnarray}
\label{PBS:curvature}
\zeta_{s}(\vec{x}) &=& \xi_{s}(\vec{x})\left(1+\frac{6}{5}\fNL\xi_{L}(\vec{x})\right) + \frac{3}{5}\fNL\xi_{s}(\vec{x})^{2}\ \\ 
\zeta_{L} &\approx& \xi_L\ .
\end{eqnarray}
Eq. \eqref{PBS:curvature} shows that the principal effect of primordial non-Gaussianity is a non-trivial correlation between the small-scale and large-scale cosmological fluctuations even at the linear level.

Considering the clustering of dark matter haloes (and more generally tracers of matter density) within a large-scale matter density fluctuation at some point $\vec{x}$, we note that the comoving number density of such objects is generically a function of the ambient large-scale matter density perturbation $(\delta_{L})$ and the magnitude of small-scale matter density fluctuations parametrized by $\sigma_{8}$, which is a functional of the small-scale linear matter density power spectrum. In the comoving synchronous gauge and at lower redshifts, one can obtain the small-scale linear matter fluctuations and the matter power spectrum in Fourier-space from the CDM+baryon transfer function
\begin{eqnarray}
\label{transfer:eq1}
\delta_{lin}(\vec{k},z) &=& T_{cb}(\vec{k},z)\zeta_{L}(\vec{k})\\
\label{eq:Pk_lin_def}
P_{lin}(k_{s},z) &=& |T_{cb}(\vec{k}_{s},z)|^{2}\zeta_{L}(\vec{k}_{s})\ .
\end{eqnarray}
where $T_{cb}(\vec{k},z)$ is the CDM+baryon transfer function computed using linear perturbation theory. The parameter $\sigma_{8}$, which characterises the magnitude of small-scale density fluctuations, can be obtained by integrating the small-scale linear matter power spectrum convolved with a suitable window function of width $8\ \text{Mpc/h}$. The window function has support over a region as small as the width of the window function ($8\ \text{Mpc/h}$) over which the large-scale component of $\zeta$ (namely $\zeta_{L}$) is nearly constant.   
Owing to the correlations between small and large-scale perturbations induced by non-Gaussian initial conditions following Eqs. \eqref{eq:Pk_lin_def} and \eqref{PBS:curvature}, $\sigma_{8}$ is now a position-dependent function following Eq. \eqref{PBS:curvature}:
\begin{eqnarray}
\label{sigm8:eq1}
    \sigma^{2}_{8}(\vec{x},z) &=& \sigma^{2}_{8,o}\left(1+\frac{6}{5}\fNL\zeta_{L}(\vec{x})\right)^{2} + \Sigma^{2}_{8}(z)\ ,
\end{eqnarray}
where $\sigma^{2}_{8,o}$ is the squared amplitude of small-scale matter density fluctuations in the absence of any primordial non-Gaussianity and is therefore only dependent on the redshift $z$. The dependence of $\sigma_{8}$ on the ambient large-scale curvature perturbation in the presence of non-Gaussian initial conditions is responsible for making the bias of dark matter density tracers scale-dependent. The position-independent term $\Sigma_{8}$ just offsets the scale-independent part of the bias, which we regard as a free parameter in our analysis. We can therefore ignore $\Sigma_{8}$ as we proceed to extract the scale-dependence of the bias following Eq. \eqref{sigm8:eq1}. In Fourier-space,
\begin{eqnarray}
\label{sigma8:eq2}
\sigma_{8}(\vec{k},z) &=& \sigma_{8,o}\left(1+\frac{6}{5}\fNL\zeta_{L}(\vec{k})\right)\ .
\end{eqnarray}
The mean comoving number density of matter density tracers $n[\delta_{L},\sigma_{8},\fNL]$ within the large-scale perturbation $\delta_{L}$ is therefore 
\begin{eqnarray}
    n[\delta_{L},\sigma_{8},\fNL] = n[0,\sigma_{8,o},\fNL]  + \frac{\partial n}{\partial\delta_{L}}[0,\sigma_{8,o},\fNL]\delta_{L}  \\ +
      \frac{\partial n}{\partial \log\sigma_{8}}[0,\sigma_{8,o},\fNL]\frac{6}{5}\fNL\zeta_{L}(\vec{k})  + ...\ ,\nonumber
\end{eqnarray}\label{eq:5}
where $\partial n/\partial \delta_{L}$ is now the scale-independent bias \textit{in the presence of} $\fNL$ and includes corrections due to the constant term in Eq. \ref{sigm8:eq1}.

The fluctuation in the (comoving) tracer number density may thus be related to the large-scale matter density perturbation as 
\begin{eqnarray}
\label{biaseq:1}
\frac{\delta n}{\overline{n}} = \left(\frac{\partial \log n}{\partial\delta_{L}}[0,\sigma_{8,0}]+\frac{\partial\log n}{\partial\log\sigma_{8}}\cdot\frac{6\fNL}{5T_{cb}(k,z)}\right)\delta_{L}\\ + ...\nonumber
\end{eqnarray}
In the above equation, $b_{L} = \partial\log n/\partial\delta_{L}$ is the (scale-independent) Lagrangian bias of tracers. 
This linear Lagrangian bias acquires a scale-dependent correction in the presence of local primordial non-Gaussianity that may be written as  
\begin{eqnarray}
\label{eq:deltabLdef}
    \Delta b_{L}(k,z) &=& 2\fNL\frac{\beta(k,z)}{\alpha(k,z)};\\
    \beta &=& \frac{\partial \log n}{\partial \log \sigma_{8}}[\delta_{L}=0,\sigma_{8,o}];\\
    \alpha(k,z) &=& \frac{5}{3}T_{cb}(k,z).
\end{eqnarray}
The functional dependence of the tracer number density $n$ on the small and large-scale density perturbations is in general highly non-trivial and may not be expressible analytically (see, e.g., Refs. \cite{Scoccimarro:2011pz, Biagetti:2016ywx}).   

The parameter $\beta = \partial\log n/\partial\sigma_8$ is essentially the same as the parameter $b_{\phi}$ in the Effective Field Theory of Large-Scale Structure (EFTofLSS)~\cite{PhysRevD.106.043506} and should, in principle, be considered independent of the lagrangian galaxy bias $(b_L)$. In the particular case of the tracer number density having the same form as a universal halo mass function, the quantity $n$ is a function only of the  peak height $\delta_{c}/\sigma_{8}$, where, for example, $\delta_{c} = 1.686$ is the threshold matter overdensity for spherical collapse in an Einstein-deSitter universe. For a tracer number density of this kind, one can show that $\beta(k,z) = b_{L}\delta_{c}$; so that the total Lagrangian linear bias in the presence of local primordial non-Gaussianity is
\begin{eqnarray}
\label{bias:final}
b_{L}(k,z;\fNL) = b_{L}(z)\left(1 + 2\fNL\frac{\delta_{c}}{\alpha(k,z)}\right)\ .
\end{eqnarray}
In this work, we perform our analysis with the universal ansatz in Eq. \eqref{bias:final} as a modelling choice.\footnote{This universal ansatz for the non-Gaussian correction to the bias has long been known to have limitations. The precise form of the non-Gaussian bias in Eq.~(\ref{eq:deltabLdef}) in terms of the Gaussian bias may have corrections, and, in particular, corrections that are tracer dependent (see, e.g., Refs.~\cite{Slosar:2008hx,Lazeyras:2022koc}). Addressing this requires introducing additional free parameters that break the relationships between terms in the bias expansion, including between the neutrino and $\fNL$ terms, that should only further degrade constraints. } 

To obtain $\alpha(k,z) = (5/3)T_{cb}(k,z)$, one can invert the Poisson equation relating the gravitational potential to the matter overdensity and write the CDM+baryon transfer function $T_{cb}(k,z)$ as 
\begin{eqnarray}
\label{transfer:alternate}
T_{cb}(k,z) = \frac{2k^2 \mathrm{T}(k)D(z)}{5H^{2}_{0}\Omega_{m}}\ ,
\end{eqnarray}
where $-(3/5)T(k)$ is the transfer function for $\phi$ deep in the matter dominated era and $D(z)$ is the linear growth factor of dark matter perturbations normalised to be the scale factor in the matter dominated era. Deep in the matter dominated era, the gravitational potential is constant with time and approaches $-(3/5)\zeta$ as $k\rightarrow 0$\footnote{In this paper, we are using a sign convention whereby the primordial curvature perturbation can be $\zeta\sim -\phi + (1/3)(\delta\rho)/(\overline{P}+\overline{\rho})$ in the Newtonian gauge and all transfer functions are defined w.r.t $\zeta$.}. That is why the function $T(k)$ is redshift independent and $T(k) \rightarrow 1$ as $k\rightarrow 0$.

\begin{figure}[h!]
    \centering
    \includegraphics[width=\columnwidth]{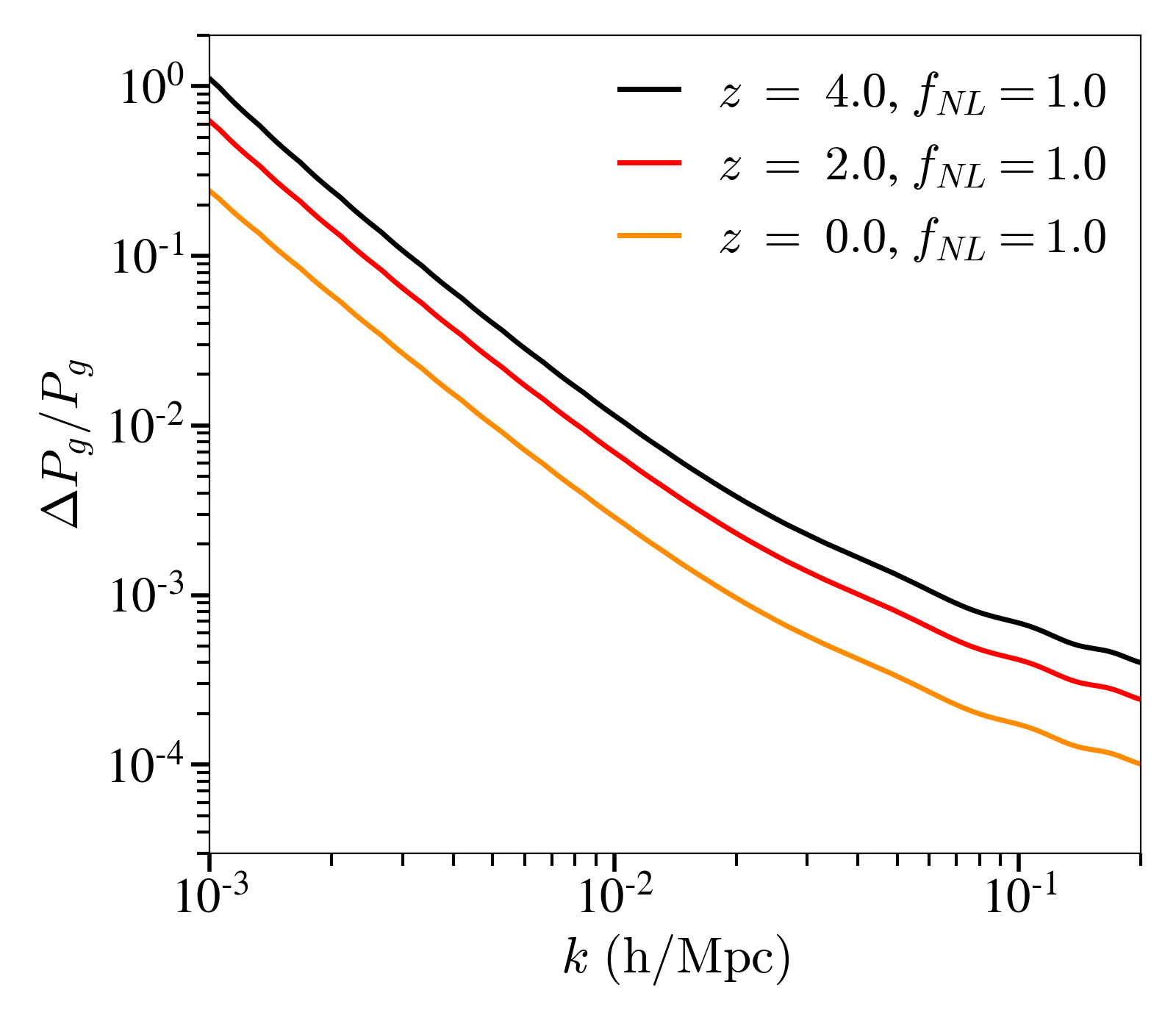}
    \caption{Fractional change in the galaxy power spectrum due to local $\fNL=1$.}
    \label{fig:fNL_scale_dependent_bias}
\end{figure}

Figure \ref{fig:fNL_scale_dependent_bias} shows the effect of local primordial non-Gaussianity on the galaxy power spectrum computed using the linear bias model for $\fNL=1$, which is an important theoretical target because $\fNL=1$ gives a natural ${\cal O}(1)$ coefficient if $\zeta$ can be expressed as a power series in a Gaussian random field and also is a typical value found in multi-field inflationary models.  We see that the small-$k$ enhancement of the galaxy power spectrum due to local primordial non-Gaussianity increases out to arbitrarily large scales. Such an effect on galaxy clustering that persists out to arbitrarily large scales can only have a primordial origin. That is why surveys like SPHEREx designed to improve measurements of galaxy clustering at large scales provide important probes into the physics of the early universe. 

\section{Non-primordial sources of galaxy clustering bias}
\label{sec:nonPbk}
Non-inflationary mechanisms having to do with the non-linear evolution of matter density fluctuations introduce their own scale-dependent features in the bias of dark matter tracers. Typically, these scale-dependent changes occur on scales comparable to the non-linear scale, which characterizes the typical distance CDM and baryons move. Yet, fast-moving particles such as photons and neutrinos can imprint changes to the bias at much lower $k$. While these late-time, causal sources of scale-dependence can only persist out to scales about as large as the horizon, they are nevertheless potentially important systematic effects that need to be considered in measurements of primordial non-Gaussianity from galaxy surveys. We consider two such sources of scale-dependent bias -- the free-streaming of neutrinos (light relics) and fluctuations in the background of ionising radiation. 

\subsection{Neutrinos}
\label{ssec:neutrinobk}
Free-streaming particles like neutrinos, which are relativistic until late cosmic times, suppress the linear growth of matter density perturbations on scales below their free-streaming scales. Moreover, the non-linear coupling between small-scale and large-scale matter density perturbations becomes dependent on the large-scale \cite{LoVerde:2014pxa, LoVerde:2016ahu, Hu:2016ssz, Chiang:2016vxa, Chiang:2017vuk, Jamieson:2018biz, Chiang:2018laa, Shiveshwarkar:2020jxr,Xu:2020fyg,Xu:2021rwg,DePorzio:2020wcz,PhysRevD.98.043503}\footnote{Note that by `scale', we refer to a comoving wavenumber.}. The linear growth factor $D(z)$ of small-scale matter perturbations depends (to leading order) on an ambient large-scale matter density perturbation $\delta_{L}(k_{L},z)$ as 
\begin{eqnarray}
\label{SU_response_def}
D(z;\delta_{L}) = D(z)\left(1+R_{D}(k_{L},z)\delta_{L}\right)\ ,
\end{eqnarray}
where we have defined $R_{D}(k_{L},z)$ as the response of the linear growth factor to an ambient large-scale matter perturbation $\delta_{L}$\footnote{We note that the separate universe approach described in this section can also be used to find the Gaussian bias $b_L$ via $b_L = d\log n/d\delta_L = d\log n/d\log\sigma_8 (d\log\sigma_8/d\delta_L)$. This gives an alternative approach to relating the Gaussian bias $b_L$ to the $\fNL$-dependent correction in Eq.~\ref{eq:deltabLdef}. Under the common assumption that $n(\sigma_8)$, this gives $\beta = b_L/(d\log\sigma_8/d\delta_L)$, which for an Einstein-deSitter Universe is $\beta = 21/13 b_L \approx 1.62 b_L$, which is remarkably similar to the $\beta = b_L\delta_{c} \approx 1.686b_L$ used in Eq.~\ref{bias:final}.}. The dependence of the growth of small-scale matter perturbations within a large-scale density perturbation on the large-scale  $k_{L}$ will naturally, following the peak-background-split argument, show up in the Lagrangian bias $b_{L}$ (used in Eq. \eqref{bias:final}) which, as a result, becomes scale-independent. Studies using the Separate Universe simulations in a $\nu\Lambda CDM$ cosmology \cite{Chiang:2017vuk,Shiveshwarkar:2020jxr}, as well as studies using large-scale simulations with neutrinos and CDM \cite{Chiang:2018laa}, show that the scale-dependence of the Lagrangian halo bias $b_{L}$ is the same as the linear growth factor response, i.e.
\begin{eqnarray}
\label{bias_scale_dependence_neutrinos}
\frac{b_{L}(k_{1},z)}{b_{L}(k_{2},z)} = \frac{R_{D}(k_1,z)}{R_{D}(k_2,z)}\ .
\end{eqnarray}

We can therefore express the Lagrangian bias $b_{L}$ (in the absence of primordial non-Gaussianity) at an arbitrary scale $k$ in terms of its value at a fixed small-scale $k_{\text{max}}$ and the \textit{relative} growth factor response as 
\begin{eqnarray}
\label{bias_neutrinos_final}
b_{L}(k,z) &=& R_{rel}(k,z)b_{L}(k_{\text{max}},z), \\
&=& \left[\frac{R_{D}(k,z)}{R_{D}(k_{\text{max}},z)}\right]b_{L}(k_{\text{max}},z)\ .\nonumber
\end{eqnarray}

\begin{figure}[h!]
    \centering
    \includegraphics[width=\columnwidth]{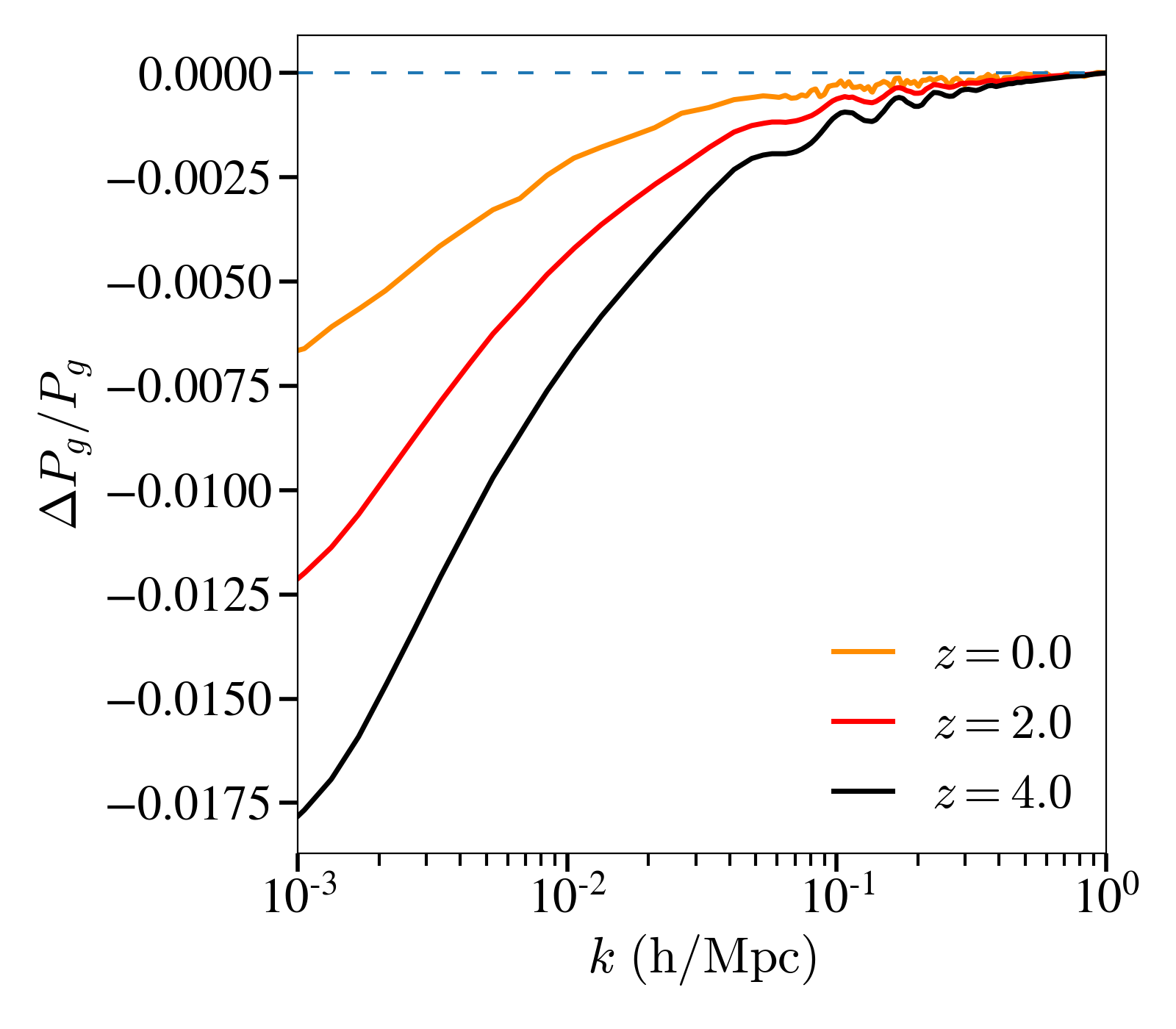}
    \caption{Fractional change in the galaxy power spectrum due to scale-dependent galaxy bias introduced by three degenerate neutrinos with total mass $M_{\nu}=0.06\ \text{eV}$. Note that the effect on the matter power spectrum due to the neutrino mass sum itself is omitted, as it is included in the reference spectrum.}
    \label{fig:dlogPg_neutrinos}
\end{figure}

\begin{figure}[h!]
    \centering
    \includegraphics[width=\columnwidth]{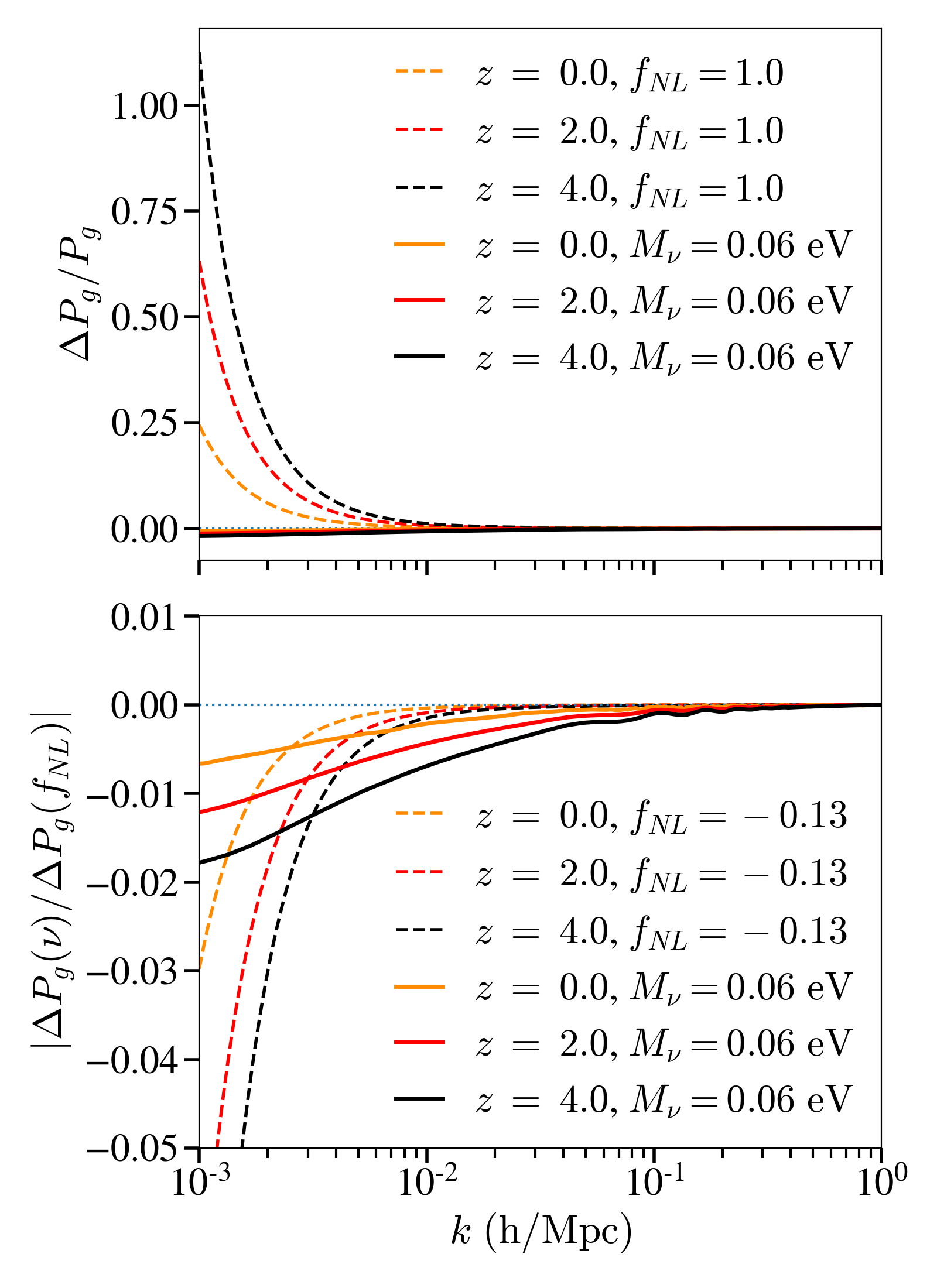}
    \caption{(Top) Effect of $\fNL=1$ and the scale-dependent galaxy bias due to three degenerate neutrinos with total mass $M_{\nu}=0.06\ \text{eV}$ on the galaxy power spectrum. (Bottom) Effect of the scale-dependent galaxy bias due to three degenerate neutrinos with total mass compared to the effect of $\fNL=-0.13$, which is about the magnitude of the bias in $\fNL$ we later obtain upon ignoring the scale-dependent bias induced by free-streaming neutrinos of the same total mass.}
    \label{fig:fNL_vs_neutrino_bias_comparison}
\end{figure}

Figure~\ref{fig:dlogPg_neutrinos} shows the scale-dependence in the galaxy power spectrum (computed using a linear galaxy bias model) introduced by the presence of three neutrinos with total neutrino mass of $M_{\nu}=0.06\ \text{eV}$, whereas figure~\ref{fig:fNL_vs_neutrino_bias_comparison} shows how this effect compares with the effect of local $\fNL=1$. The overall amplitude of the scale-dependent bias from neutrinos scales roughly with $M_\nu$~\cite{Shiveshwarkar:2020jxr}. Figure \ref{fig:dlogPg_neutrinos} shows that this effect increases with redshift and, for $M_\nu = 0.06$ eV, is no more than $2\%$ even at the highest redshifts probed by SPHEREx ($z_{\text{max}}\approx 4.3$). Figure \ref{fig:fNL_vs_neutrino_bias_comparison} also shows that while this effect is very small compared to the effect of local primordial non-Gaussianity at the largest scales -- it is comparable to the effect of a small local primordial non-Gaussianity of $(\fNL= -0.13)$ at intermediate scales ($k\approx 0.01\ \text{h/Mpc}$). Figure \ref{fig:fNL_vs_neutrino_bias_comparison} shows that the shapes of the neutrino and $f_{\rm NL}$ scale-dependent biases are quite different; so it may be surprising that there is any degeneracy between the two signals. Indeed, we find (as we shall show later) that the existence of the neutrino-induced scale-dependent bias does not significantly alter the MCMC as well as Fisher forecasts for $f_{\rm NL}$, as long as the effect of neutrinos is modelled appropriately. However, the fact that the neutrino effect is comparable in magnitude to the effect of a small local primordial non-Gaussianity (as is shown in figure \ref{fig:fNL_vs_neutrino_bias_comparison}) over the range of scales probed by a galaxy survey\footnote{Note that galaxy surveys measure galaxy power spectra at finite scales limited by the survey volume ($k\gtrsim V^{-1/3}$).} such as SPHEREx leads to a small but significant bias in the MCMC forecast of $f_{\rm NL}$ if the aforementioned effect is not included in theoretical models of the observed galaxy power spectra.

\subsection{Fluctuations in the background of ionising radiation}
\label{ssec:UVbk}
A background of ionising radiation pervades the Cosmos after the epoch of reionization and this radiation affects how gas cools and forms galaxies. Fluctuations in the density of ionising radiation in turn modulate the clustering of galaxies in a manner that has a scale-dependent bias on scales smaller than the mean free path for ionising photons to be absorbed or, when the mean free path is long, the distance the photons are able to travel since being emitted. Indeed, the mean free path is long, approaching the comoving Hubble radius by $z\sim 2$.  Thus, ionizing radiation can potentially imprint an additional scale-dependence in the galaxy power spectrum \cite{Sanderbeck:2018lwc}.  One way to incorporate this effect is to realise that fluctuations in the number density of a certain population of galaxies (negatively) trace the underlying fluctuations of ionising radiation (in addition to matter) with a corresponding \textit{intensity} bias $b_{J}$:
\begin{eqnarray}
\delta_{g} &=& b_{g}\delta_{c}-b_{J}\delta_{J}\ .
\label{eq:dg_with_bJ}
\end{eqnarray}
In other words, intensity bias $b_{J}$ encodes how the clustering of a galaxy sample responds to ambient fluctuations in the ionising radiation background \cite{Sanderbeck:2018lwc}. In Eq. \eqref{eq:dg_with_bJ} and in all equations hence, $\delta_{c}$ refers to the density fluctuations of Cold Dark Matter (CDM) and baryons. Similarly, in what follows, we use the term `matter' to mean CDM+Baryons.

The effect of ionising radiation fluctuations on galaxy power spectra is two-fold: there is a scale-dependent suppression of power as large as the cross-power spectrum between matter and ionising radiation ($P_{cJ})$ and there is an additional shot noise contribution due to the discreteness of sources of UV fluctuations. 
Assuming that $P_{cJ}=\langle\delta_{J}^{\dag}\delta_{c}\rangle$ is real\footnote{As there is no preferred direction in the universe, this two-point function is a Fourier transform of a radial function which is always real. Time/Redshift is a preferred direction that slightly breaks this.},
we have the following expressions for the galaxy power spectra and the covariance matrix of galaxy number density contrasts:
\begin{eqnarray}
\label{Pg_with_bJ}
P_{g} = P_{cc}\left(b_{g}-b_{J}\frac{P_{cJ}}{P_{cc}}\right)^{2}+b_{J}^{2}P_{Jshot}.
\end{eqnarray}
In the above equation, $P_{cc}$ refers to the power spectrum of CDM+Baryon density fluctuations\footnote{Note that we \textit{do not} include the contribution of massive neutrinos when we refer to $P_{cc}$ as the matter power spectrum. N-body simulations performed in~\cite{Castorina2014} suggest that halo bias defined w.r.t CDM+Baryon density perturbations is likely to be more universal. On the other had, a galaxy bias defined w.r.t the total matter fluctuation including the contribution of massive neutrinos can introduce an unphysical scale-dependence in the galaxy bias ~\cite{2014JCAP02049C,Raccanelli:2017kht,Vagnozzi:2018pwo}. }, whereas $P_{cJ}$ refers to the cross-power spectrum of matter and ionising radiation fluctuations. $P_{Jshot}$ is the shot noise contribution that arises due to the discreteness of sources of UV fluctuations (i.e. quasars). The minus sign in Eq. \eqref{eq:dg_with_bJ} and Eq. \eqref{Pg_with_bJ} encodes the general fact that excess ionising radiation typically decreases the rate of galaxy formation and reduces the observed number density of galaxies. 

Modelling the intensity bias is a non-trivial task, as it depends on the complicated physics of galaxy formation in dark matter haloes. This presents an additional challenge, as $b_J$ should be a population and redshift-dependent parameter. Ref.~\cite{Sanderbeck:2018lwc} calculated how the cooling rate for the gas in dark matter haloes is affected by fluctuations in the ionizing background and found that the cooling rate changes by a factor of $\sim 0.1$ in response to a fluctuation in the background for relevant densities and temperatures of virialized gas of galaxy-hosting dark matter haloes.  In a model where the cooling rate of this gas is directly tied to the observed properties of a galaxy (like its star formation rate), this would lead to $b_J \sim 0.1$.  However, there are likely to be feedback effects that reduce this response.  Thus, they estimated that $b_{J}$ for typical galaxy samples is at most $0.1$ and could be much smaller. A reasonable value for $b_{J}$ is therefore somewhere between $0$ and $0.1$. For simplicity, we assume a fixed value of $b_J = 0.05$ for all galaxy samples and redshifts unless otherwise specified. 

The shape of the matter-UV background cross-power spectrum $(P_{cJ})$, UV auto-power spectrum $(P_{JJ})$, as well as the shot noise contribution $(P_{Jshot})$ should be more straightforward to model. The $k$-dependence of the UV transfer function $(T_{J} = P_{cJ}/P_{cc})$ and the shot noise $(P_{Jshot})$ arises from the finite mean free path of UV photons: $T_{J}^{2}$ and $P_{Jshot}$ decay as $k^{-2}$ at scales smaller than the UV photon mean free path~\cite{Sanderbeck:2018lwc}. The shape of the matter-ionising radiation cross-power spectrum is relatively unchanged with cosmology~\cite{Sanderbeck:2018lwc} and we use the template provided by Ref.~\cite{Sanderbeck:2018lwc} in our analysis. On the other hand, the shot noise power varies inversely as the effective number density of ionising radiation sources~\cite{Sanderbeck:2018lwc}; it tends to be large (and increase with redshift) because relatively rare systems -- quasars -- contribute most or all of the background \cite{Haardt:2011xv}. In addition, the extent of the shot noise depends on the lifetime $(t_{Q})$ of ionising radiation sources (i.e. quasars) -- with an infinite lifetime model for quasars in which they are assumed to contribute 100\% of the background giving a maximal estimate of the shot noise~\cite{Sanderbeck:2018lwc}. 

\begin{figure}[h!]
    \centering
    \includegraphics[width=\columnwidth]{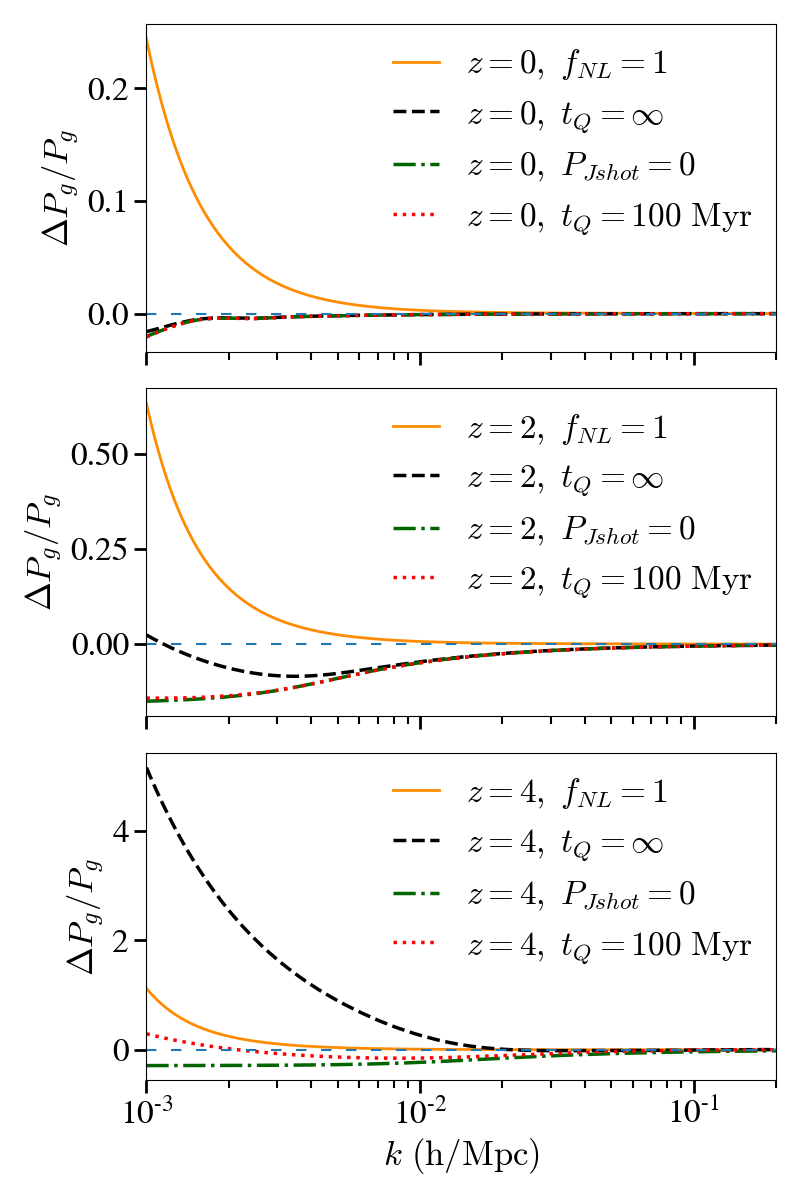}
    \caption{Effect of ionising radiation fluctuations with $b_{J}=0.05$ in comparison to the effect of $\fNL=1$ on the galaxy power spectrum for different models of the quasar shot noise. The effect of ionising radiation is dominated by the shot noise of ionising radiation sources at higher redshifts for the extreme case of near-infinite quasar lifetimes.}
    \label{fig:dlogPg_bJ_comparison}
\end{figure}

Figure \ref{fig:dlogPg_bJ_comparison} shows the fractional effect of ionising radiation on the galaxy power spectrum in comparison to the effect of scale-dependent bias due to $\fNL=1$ for different models of the quasar shot noise at the redshifts relevant for SPHEREx.  Here $b_J = 0.05$ is assumed, near the values estimated in the simple model in \citet{Sanderbeck:2018lwc}, which should be thought of as a rough upper bound on the effect.  The matter-density tracing component of the ionising radiation fluctuations given by $T_{J}$ causes a significant suppression of power out to scales comparable to the horizon. This effect increases with redshift until the shot noise term in Eq. \eqref{Pg_with_bJ} becomes large enough to lead to a relative increase in power at larger scales. Thus, both the quasar shot noise and the density tracing component of ionising radiation fluctuations give rise to features in the galaxy power spectrum that become important at near-horizon scales. Figure \ref{fig:dlogPg_bJ_comparison} shows that for the most reasonable upper limit on quasar lifetime, which is $100\, \text{Myr}$, the effect of shot noise is subdominant to that of the matter-density tracing component of the ionising radiation fluctuations except at higher redshifts ($z\approx4-5$), at which point the two effects become comparable to each other. For the extreme case of infinite quasar lifetime, the quasar shot noise dominates over the clustered component at larger scales, especially at the highest redshifts ($z\approx 4-5)$.\footnote{Of course, the ionising radiation contributions stop increasing at much larger scales ($k\lesssim 0.0005\ \text{h/Mpc}$) so the bias eventually becomes constant at sufficiently low $k$. }

\section{Galaxy power spectrum model}
\label{sec:Pk}
We model the galaxy power spectra according to the linear bias model, whereby the galaxy number density contrast is proportional to the CDM+baryon overdensity, $\delta_{g}\propto\delta_{c}$ and the proportionality constant is the \textit{galaxy} bias. However, peculiar velocities of galaxies modulate the galaxy power spectra in redshift-space through redshift-space distortion. This is a line-of-sight effect that breaks the isotropy of the matter power spectrum and causes the galaxy power spectrum to only by isotropic around the line of sight. As a consequence, the galaxy power spectrum in Fourier-space depends on the magnitude of the wave vector $k$ and the angle between the wave vector and the line of sight encoded in its cosine $\mu = k_{||}/k = \vec{k} \cdot \vec{r}/(kr)$. At linear and mildly non-linear scales, the redshift-space distortion is given by the Kaiser formula~\cite{Kaiser:1987qv}, whereby the galaxy power spectrum for a given sample of galaxies is 
\begin{eqnarray}
\label{Kaiser_formula}
P_{g,j}(k,z) = b^{2}(k,z)\left[1+\beta_{j}^{\text{rsd}}(k,z)\mu^{2}\right]^{2}P^{L}_{cc}(k,z)\ ,
\end{eqnarray}
where $b(k,z)$ is the linear Eulerian bias of the population of galaxies generically dependent on the scale $k$ and the redshift, while $\beta_{j}^{\text{rsd}}$ is related to the linear growth rate of matter density perturbations as $\beta^{\text{rsd}}_{j} = b(k,z)^{-1} d\log\delta_{c}/d\log a$.\footnote{In the presence of free-streaming neutrinos, this term is actually going to depend on scale -- it will be suppressed at smaller scales simply because free-streaming of neutrinos suppresses the growth of matter perturbations below the free-streaming scale. We ignore this scale-dependence for the purposes of this work, because we find that its effect on forecasts for $\fNL$ is $\lesssim\ 5-6\%$ and is subleading to the effects we investigate here. } Note that $P^{L}_{cc}$ is the linear power spectrum of CDM+baryon density perturbations. However, the signal for local primordial non-Gaussianity that we are looking for -- the scale-dependent bias -- comes mainly from scales large enough to be in the linear regime.\footnote{Indeed, as we will show later, scales with $k\lesssim 0.02\ \text{h/Mpc}$ provide most of the constraining power for $\fNL$ in a measurement based on the galaxy power spectrum: $\sigma(\fNL|k\lesssim 0.02\ \text{h/Mpc})=1.004$ for SPHEREx, which is about $90\%$ of the constraining power of all the scales.} For this reason, we expect that using the linear power spectrum at small scales instead of the more appropriate non-linear power spectrum does not significantly alter the galaxy power spectrum forecast for $\fNL$.

Galaxy surveys measure the redshift and the angular position of different galaxies and convert this data into a Fourier-space galaxy power spectrum. In doing so, one needs to assume a fiducial cosmology to convert angular position and redshift into comoving Fourier-space coordinates. However, this does not yield the true galaxy power spectrum in the Fourier-space as comoving wavenumbers used in such an analysis are dependent on the fiducial cosmology. This error in the galaxy power spectrum extracted from the data but assuming a fiducial cosmology different from the true cosmology is called the Alcock-Paczynski effect. One can obtain the galaxy power spectrum in the true cosmology by mapping the comoving wavenumbers of the fiducial and true cosmologies in a way that preserves the redshift and the angular coordinates of the observed galaxy samples. This mapping is given by 
\begin{eqnarray}
\label{AP_map_1}
k_{||} &=& \frac{H(z)}{H_{\text{fid}}(z)}k^{\text{fid}}_{||}\ , \\
\label{AP_map_2}
k_{\perp} &=& \frac{D^{\text{fid}}_{A}(z)}{D_{A}(z)}k^{\text{fid}}_{\perp}\ ,
\end{eqnarray}
where $D_{A}(z)$ and $H(z)$ are the angular diameter distance and the Hubble rate in the two different cosmologies.\footnote{Note that at $z=0$, $D_{A}(z=0)=0$ and $k = k_{||} = H(z)/H_{\text{fid}}(z)\times k_{\text{fid}} $.} The mapping in Eqs. \eqref{AP_map_1} and \eqref{AP_map_2} is a mapping between the fiducial and the true cosmologies that allows one to express the observed galaxy power spectrum (in the fiducial cosmology) in terms of the true galaxy power spectrum through a redshift dependent multiplicative factor 
\begin{eqnarray}
\label{f_AP}
f_{AP}(z) = \frac{H(z)D^{\text{fid}}_{A}(z)^{2}}{H^{\text{fid}}(z)D_{A}(z)^{2}}.
\end{eqnarray}
For SPHEREx galaxy data, an uncertainty in the redshift of galaxy samples suppresses the galaxy power spectrum in the radial direction. This effect can be modelled by adding an exponential damping factor that suppresses the Fourier-space galaxy power spectrum at large $k$ along the line of sight (see, e.g., Ref.~\cite{Dore:2014cca}),
\begin{eqnarray}
\label{redshift_error_term}
f_{\sigma_{z}} = \exp\left(-k^{2}\mu^{2}\tilde{\sigma}_{z}^{2}(1+z)^{2}/H(z)^{2}\right)\ .
\end{eqnarray}
Here $\tilde{\sigma}_{z}(1+z)$ is the redshift uncertainty of a given galaxy sample. This uncertainty is larger for SPHEREx than many other spectroscopic galaxy surveys owing to its low resolution. 

There is also an additional damping factor that needs to be included to model the smearing of (small-scale) BAO features in the galaxy power spectrum due to the effect of non-linear bulk flows~\cite{Dore:2014cca}. Following Ref. \cite{Dore:2014cca}, this term is given by 
\begin{eqnarray}
\label{bulk_flows}
f_{BF}(k,z) = \exp\left(-\frac{1}{2}k^{2}\Sigma_{\perp}^{2}-\frac{1}{2}k^{2}\mu^{2}(\Sigma_{||}^{2}-\Sigma_{\perp}^{2})\right)\ .
\end{eqnarray}
 This non-linear smearing arises due to the bulk motion of dark matter particles over Lagrangian scales as large as $10\ \text{Mpc/h}$ and does not invalidate the linear bias assumption on BAO scales, as it is a resummation of the effect of large-scale linear modes that advects any large-scale structure tracer.
The Lagrangian displacement fields are given as \cite{Dore:2014cca}
\begin{eqnarray}
\Sigma_{\perp}(z) &=& c_{\text{rec}}D(z)\Sigma_{0}\ ,\\
\Sigma_{||}(z) &=& c_{\text{rec}}D(z)(1+f(z))\Sigma_{0}\ ,
\end{eqnarray}
where $D(z)$ is the linear growth factor and $f(z)$ is the linear growth rate. We set the parameter $c_{\text{rec}}=0.5$~\cite{Dore:2014cca} and $\Sigma_{0} = 11\ \text{h/Mpc}$ for a fiducial $\sigma_{8} = 0.8$.
Putting all factors together, we model the observed power spectrum of a given galaxy sample (in the absence of the effect of ionising radiation fluctuations) indexed by $j$ as 
\begin{eqnarray}
\label{P_th_model_no_bJ}
P^{th}_{g0,j}(k_{\text{fid}},\mu_{\text{fid}},z) = f_{BF}(k,\mu,z)\times f_{\sigma_{z},j}(k,\mu,z)\\
\times f_{AP}(z)\times b^{2}_{j}(k,z)\times \left[1+\beta_{j}(k,z)\mu^{2}\right]^{2}\times P^{L}_{cc}(k,z)\nonumber\ ,
\end{eqnarray}
with the wavenumber $k$ and $\mu=k_{||}/k$ on the right-hand side expressed in terms of their counterparts in the fiducial cosmology according to Eqs. \eqref{AP_map_1} and \eqref{AP_map_2}. Our model for the effect of ionising radiation according to Eq. \eqref{Pg_with_bJ} implies that in the presence of ionising radiation, 
\begin{eqnarray}
\label{P_th_model_final}
P^{th}_{g,i} = P^{th}_{g0,i}\left(1-\frac{b_{J,i}}{b_{i}}\frac{P_{cJ}}{P_{cc}}\left(1+\beta_{j}\mu^{2}\right)^{-1}\right)^{2} \\ \nonumber
+ \left(f_{AP}\times f_{BF}\times f_{\sigma_{z},i}\times b_{J,i}^{2}P_{Jshot}\right)\ .
\end{eqnarray}
The galaxy bias at arbitrary scales and redshifts includes the scale-dependent biases induced by local primordial non-Gaussianity as well as the free-streaming of neutrinos. Following Eqs. \eqref{bias:final} and \eqref{bias_neutrinos_final} the total (Eulerian) bias of a given galaxy sample is given by:
\begin{eqnarray}
\label{bias_prescription}
b_{j}(k,z) =1+(b_{j0}-1)\left[\frac{R_{D}(k,z)}{R_{D}(k_{\text{max}},z)}\right]\\
\times\left[1+\frac{2\fNL\delta_{c}}{\alpha(k,z)}\right]\ . \nonumber
\end{eqnarray}
We note that while there are additional $k$-dependent terms in the galaxy bias that grow with increasing $k$ (e.g. terms proportional to $k^2$ \cite{Chiang:2018laa}) these terms are less important in the low-$k$ regime where our non-Gaussian signal dominates, and where the neutrino and UV bias effects are important, so we neglect them for simplicity.

\section{SPHEREx survey specifications}
\label{sec:SPHEREx_survey}

SPHEREx is a proposed all-sky survey designed to reduce systematics in the measurement of galaxy distributions on large scales (between $k_{\text{min}} \approx 0.001\ \text{h/Mpc}$ and $ k_{\text{max}} \approx 0.2\ \text{h/Mpc}$)~\cite{Dore:2014cca}. It will map a large cosmic volume in 3D and will this obtain data from many more modes than CMB experiments like Planck~\cite{Planck:2018vyg,Planck:2019kim} or WMAP~\cite{WMAP:2012nax}. SPHEREx will measure the spectroscopic redshift of pre-determined galaxy populations belonging to the all-sky catalogues of the WISE~\cite{WISE}, Pan-STARRS~\cite{PanSTARRS} and DES~\cite{DES} surveys~\cite{Dore:2014cca}. The observational pipeline of SPHEREx produces a galaxy type, galaxy redshift and a redshift uncertainty. The galaxies are divided into different galaxy samples  with different galaxy biases classified according to their redshift uncertainty. The observation of different galaxy samples allows one to use multitracer techniques, which significantly improve forecasts for the measurement of primordial non-Gaussianity \cite{Yamauchi:2014ioa,Dore:2014cca}.
\begin{table}[]
    \centering
    \begin{tabular}{||c|c||} 
    \hline
    $ \text{sample} $ & $\tilde{\sigma}_{z} $ \\ [0.5ex] 
    \hline
    1 & 0.003 \\ [0.5ex]
    \hline
    2 & 0.01 \\ [0.5ex]
    \hline
    3 & 0.03 \\ [0.5ex]
    \hline
    4 & 0.1 \\ [0.5ex]
    \hline 
    5 & 0.2 \\ [0.5ex]
    \hline
    \end{tabular}
    \caption{SPHEREx galaxy samples and their redshift uncertainties. The right column shows the \textit{maximum} redshift errors for the galaxy samples listed in the first column -- for our forecasts, we assume the maximum redshift uncertainties for each galaxy in a given sample.}
    \label{SPHEREx_samples}
\end{table}

The main observables that can be obtained from the SPHEREx data are the redshift-space galaxy power spectra and the galaxy bispectra. We consider the galaxy power spectra in this paper. SPHEREx will divide its galaxy population into 11 redshift bins $\left[z^{min}_{i},z^{max}_{i}\right]^{i=11}_{i=1}$ labelled by their mean redshifts $\overline{z}_{i}$, with galaxies in each redshift bin further classified into five different galaxy samples with different redshift uncertainties $\tilde{\sigma}_{z,j}$~\cite{Dore:2014cca} (labelled samples 1 through 5 in increasing order of $\tilde{\sigma}_{z}$ as shown in table \ref{SPHEREx_samples}). Galaxies in each redshift bin have correspondingly different biases $b_{j}(k,z)$.\footnote{Details can be found at \hyperlink{https://github.com/SPHEREx/Public-products}{https://github.com/SPHEREx/Public-products} in  \hyperlink{https://github.com/SPHEREx/Public-products/blob/master/galaxy_density_v28_base_cbe.txt}{galaxy\_density\_v28\_base\_cbe.txt}} We marginalise over the small-scale galaxy biases ($b_{j0}=b_{j}(k_{\text{max}},z$)) as free parameters and express the galaxy bias at arbitrary scales and redshifts following Eq. \eqref{bias_prescription}.

\begin{figure}[h!]
    \centering
    \includegraphics[width=\columnwidth]{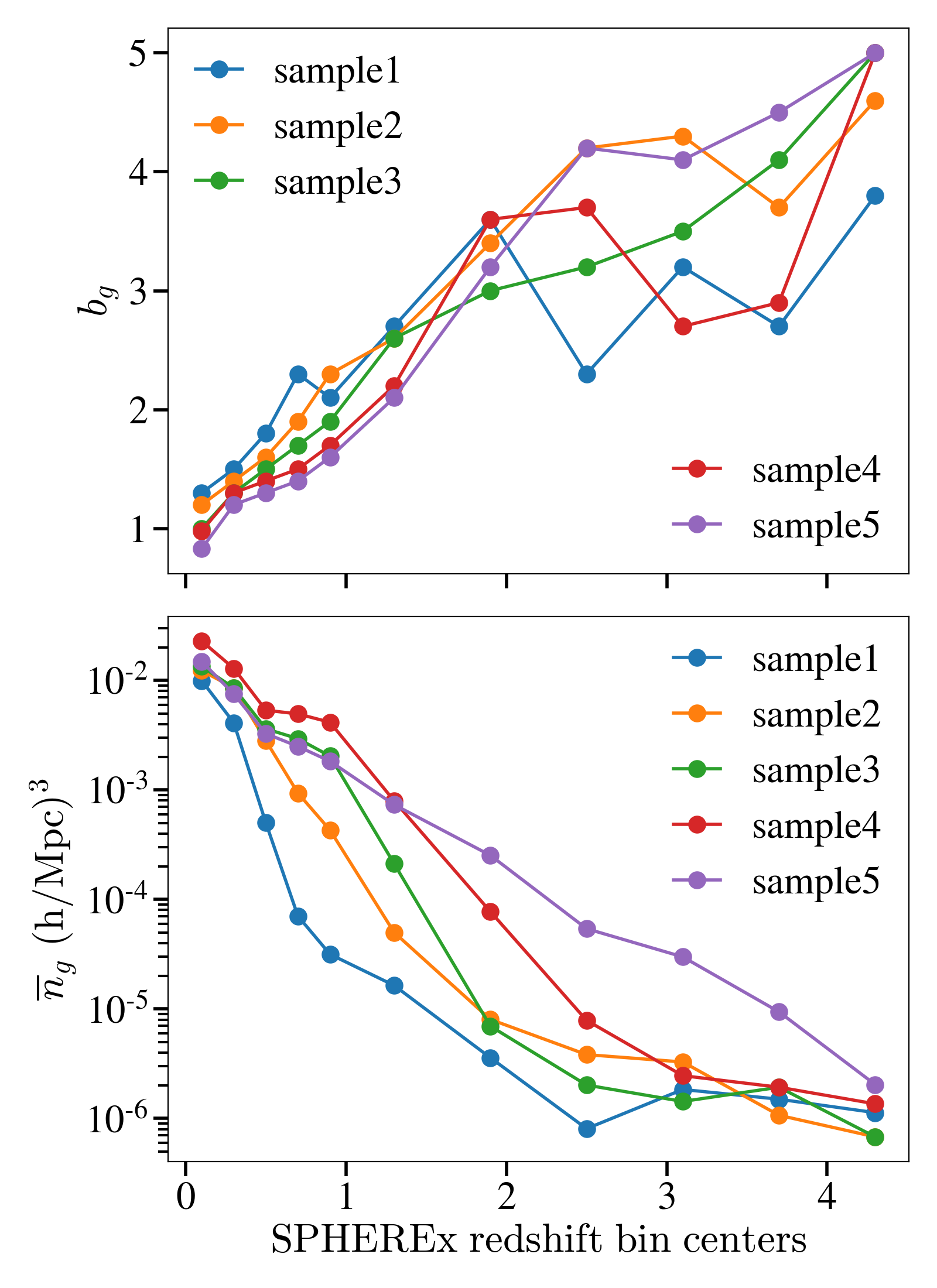}
    \caption{Galaxy number densities and galaxy biases (in the small-scale limit) for all five samples of SPHEREx labelled in increasing order of redshift uncertainty $\tilde{\sigma}_{z}$ according to table \ref{SPHEREx_samples}.}
    \label{fig:SPHEREx_bg_ng}
\end{figure}

Figure \ref{fig:SPHEREx_bg_ng} shows the number densities and biases (in the small-scale limit) for the five galaxy populations observed by the SPHEREx survey. From figure \ref{fig:SPHEREx_bg_ng}, we note that the galaxy sample 5 with the highest redshift error $(\tilde{\sigma}_{z}=0.2)$ has the highest number density at most redshifts of interest (especially at higher redshifts when the effect of $\fNL$ is most pronounced) and can be considered a \textit{high-density} sample. Analogously, we consider sample 1 $(\tilde{\sigma}_{z})$ as a \textit{low-density} galaxy sample.

\section{Galaxy likelihood for SPHEREx MCMC forecast}
\label{sec:likelihood}
We perform Markov Chain Monte Carlo (MCMC) analyses using the public code MontePython v3.4~\cite{Audren:2012wb,MontePython3} connected to a modified version of the Boltzmann code CLASS v3.0.1\footnote{We use the publicly available version of CLASS v3.0.1 including the neutrino growth response calculation found at \hyperlink{https://github.com/dsjamieson/class_public/tree/su_response_3.0}{https://github.com/dsjamieson/class\_public/tree/su\_response\_3.0}}~\cite{Blas:2011rf,Lesgourgues:2011re,Lesgourgues:2011rh} modified to compute the neutrino growth response. We construct a new galaxy power spectrum likelihood within MontePython v3.4 for the SPHEREx survey with five galaxy samples and 11 redshift bins by generalising the Euclid galaxy power spectrum likelihood~\cite{Sprenger:2018tdb} and updating it to the SPHEREx specifications (see section~\ref{sec:SPHEREx_survey}), plus adding the new ingredients to the galaxy power spectrum (see section~\ref{sec:Pk} for primordial non-Gaussianity (see section~\ref{sec:fNLbias}, neutrino growth response (see section~\ref{ssec:neutrinobk}) and UV-background radiation (see section~\ref{ssec:UVbk}).

We conduct our MCMC runs for the following set of cosmological parameters : \{${\omega_{b}, \omega_{cdm}, 100\theta_s,A_s,n_s,z_{reio}, M_{\nu}, \fNL}$\}. 
Additionally, we regard the scale-independent parts of the galaxy bias ($=1+b_{L}(k_{\text{max}},z)$ in Eq. \eqref{bias_neutrinos_final}) as well as the bulk flow parameter $\Sigma_{0}$ as free parameters. We marginalise over these 56 additional nuisance parameters to obtain forecasts for $\fNL$. To break parameter degeneracies, we include a mock CMB power spectrum likelihood named `fake\_planck\_realistic' in MontePython from Ref.~\cite{Brinckmann:2018owf} (see that reference for a detailed description of this likelihood designed to mimic the sensitivity of Planck 2018)\footnote{Note that our purpose in including a mock CMB likelihood based on Planck 2018 is to resolve parameter degeneracies and is essentially equivalent to imposing Planck-based priors on our cosmological parameters as is done in~\cite{Dore:2014cca}. By the time SPHEREx data is released, the then-current CMB dataset would likely have been collected by CMB-S4~\cite{Abazajian:2019eic}, which is projected to be more constraining than the Planck datasets}. We use the option to not include the CMB lensing power spectrum so the mock CMB likelihood is not significantly correlated with our SPHEREx galaxy power spectrum likelihood.\footnote{Although there is still lensing in the CMB temperature and polarization power spectra, we assume it is safe to neglect the correlation with matter density at angular scales $l\lesssim 3000$ at Planck-level precision.} In order to isolate the ability of SPHEREx to constrain primordial non-Gaussianity, we have not added the effect of primordial non-Gaussianity to the mock CMB likelihood and so it does not directly carry information about $\fNL$. We use the modified version of CLASS to obtain matter power spectra and the neutrino growth response function $R_{D}(k, z)$ defined in Eq. \eqref{SU_response_def}. For the matter-ionising radiation cross power spectrum and the quasar shot noise, we use the templates computed in Ref. \cite{Sanderbeck:2018lwc}.

To construct the SPHEREx likelihood, we work with the standard assumption \cite{Dore:2014cca} that the galaxy density field $\delta_{g,i}$ of any galaxy sample at a given (fiducial) $\textbf{k}$-mode and redshift is a Gaussian random variable and that galaxy densities at different redshifts are uncorrelated. This results in an \textit{exponential} likelihood for the galaxy power spectra (at a given $\textbf{k}$-mode and redshift) that goes as 
\begin{eqnarray}
    \mathcal{L}_{\textbf{k},z} = \frac{1}{\pi^{N}\det{\mathcal{C}}}\cdot\exp\left[-\text{Tr}\left(\mathcal{C}^{-1}\mathcal{D}\right)\right]\ , \label{galaxy_likelihood_form}
\end{eqnarray}
up to a multiplicative factor that does not depend on the sampled cosmological parameters. Here $\mathcal{C}$ is the covariance matrix of galaxy number densities and depends on the sampled cosmological and nuisance parameters in an MCMC run:
\begin{eqnarray}
\label{eq:data_cov_expression}
\mathcal{C}_{ij} = \langle\delta_{i}^{\dagger}\delta_{j}\rangle =  P^{th}_{ij} + \delta^{K}_{ij}\frac{1}{\overline{n}_{i}}\ ,
\end{eqnarray}
where $P^{th}_{ij}$ is the (observed) cross-power spectrum of the $i^{\text{th}}$ and $j^{\text{th}}$ galaxy populations of SPHEREx and $\overline{n}_{i}$ is the mean number density of galaxies in the $i^{\text{th}}$ sample. 

On the other hand, the matrix $\mathcal{D}$ depends on the \textit{fiducial} cosmological (table \ref{fiducial_cosmology}) and nuisance parameters. 
\begin{eqnarray}
\label{Data_matrix}
\mathcal{D}_{ij} = \langle\delta_{i}^{\dagger}\delta_{j}\rangle_{fid} = P^{fid}_{ij} + \delta^{K}_{ij}\frac{1}{\overline{n}_{i}}\ ,
\end{eqnarray}
where $P^{fid}_{ij}$ is the (fiducial) cross-power spectrum of the $i^{\text{th}}$ and $j^{\text{th}}$ galaxy samples of SPHEREx. We present the complete expressions for the fiducial and theoretical cross-power spectra in Appendix (A). 

The entire SPHEREx galaxy power spectrum likelihood is just the product of $\mathcal{L}_{\textbf{k},z}$ over all the 11 redshift bins and independent \textbf{k}-modes between $k_{\text{min}} = 0.001\ \text{h/Mpc}$ and $k_{\text{max}} = 0.2\ \text{h/Mpc}$. In taking this product, we are assuming that the likelihoods $\mathcal{L}_{\textbf{k},z}$ at different redshifts and $\textbf{k}$ are independent of each other. This is a widely used approximation while doing MCMC and Fisher forecasts which avoids complicated windowing effects~\citep{Dore:2014cca,Sprenger:2018tdb}. An MCMC analysis used for cosmological forecasts typically maximises the likelihood by minimising~\cite{Sprenger:2018tdb} \[ \chi^{2} = -2\sum_{z_{i}}V(z_{i})\int_{k_{\text{min}}}^{k_{\text{max}}}\int_{-1}^{1}\frac{k^{2}dkd\mu}{2(2\pi^{2})}\log\mathcal{L}_{\textbf{k},z}\ ,\] where $V(z_{i})$ is the comoving volume in the $i^{\text{th}}$ redshift bin.

\section{Results}
\label{sec:results}
 \begin{figure}[h!]
    \centering
    \includegraphics[width=\columnwidth]{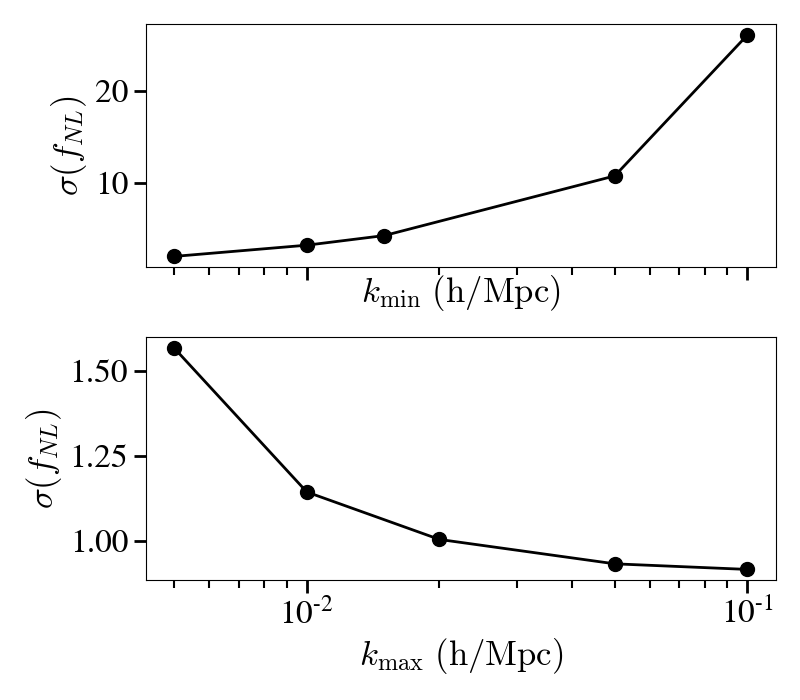}
    \caption{(Top) Fisher forecast of $\fNL$ obtained from the SPHEREx likelihood with variable $k_{\text{min}}$ and $k_{\text{max}}=0.2\ \text{h/Mpc}$ (Bottom) Fisher forecast of $\fNL$ obtained from the SPHEREx likelihood with variable $k_{\text{max}}$ and $k_{\text{min}}=0.001\ \text{h/Mpc}$.}
    \label{fig:sig_fNL_kmin_kmax}
\end{figure}
To study the impact of neutrinos and ionising radiation on future $\fNL$ measurements we perform three categories of forecasts. First, we conduct forecasts on $\fNL$ in the absence of these effects. Second, we perform forecasts including the effects of neutrinos and ionising radiation in our fiducial galaxy power spectra, but neglecting them in the theoretical galaxy power spectra entering into the covariance matrix. This allows us to estimate the bias in $\fNL$ if these effects are neglected. Third, we include the new sources of scale-dependent bias in both fiducial and theoretical galaxy power spectra. By comparing our first and third forecasts, we learn how neutrinos and ionizing radiation change the forecasted sensitivity and bias for $\fNL$. Throughout our analysis we find excellent agreement between the results obtained by MCMC analysis and from Fisher matrix calculations. 

In the absence of the effect of neutrinos and ionising radiation on the large-scale galaxy power spectra, we obtain an MCMC forecast of $\sigma(\fNL)=0.91$ around $\fNL=1.0$. This is consistent with the Fisher matrix analysis in Ref. \cite{Dore:2014cca}, who found $\sigma(\fNL)=0.87$, and more recent SPHEREx forecasts \cite{Chen}. This forecast is obtained from the galaxy power spectrum on scales with $k_{\text{max}}=0.2\ \text{h/Mpc}$ and $k_{\text{min}} = 0.001\ \text{h/Mpc}$. This range of scales would, in general, be different for different galaxy surveys. Figure \ref{fig:sig_fNL_kmin_kmax} shows how the SPHEREx $\fNL$ Fisher forecast depends on the largest or smallest scales in the galaxy survey. From figure \ref{fig:sig_fNL_kmin_kmax}, we note that $\sigma(\fNL)\approx 1.00$ for $k_{\text{max}}=0.02\ \text{h/Mpc}$. This means that the bulk of our forecast on $\fNL$ comes from relatively large-scales $k\lesssim 0.02\ \text{h/Mpc}$, thereby justifying our use of the linear power spectrum for this analysis\footnote{When non-linear scales have more significant constraining power (as when constraining non-local $\fNL$ for example), the linear modelling used in this paper is not adequate -- one needs to use a more accurate and comprehensive framework such as the Effective Field Theory of Large-Scale Structure (EFTofLSS). See~\cite{Carrasco:2012cv} for more details. The non-linear terms in the galaxy bias expansion according to the EFTofLSS encode small-scale physics that, to a good approximation, can be ignored at the relatively large scales which dominate our forecast for $\fNL$.}. Our forecast for $n_{s}$ ($\sigma(n_{s})=2.5\times 10^{-3}$) is also consistent with the Fisher matrix analysis of Ref. \cite{Dore:2014cca}.

In general, the constraining power of a given galaxy population would also depend on factors specific to the galaxy population -- namely the galaxy bias, number density and (in case of SPHEREx) the redshift uncertainty $\tilde{\sigma}_{z}$. The higher the number density, the lower the shot noise and the better the forecast whereas a larger redshift uncertainty significantly limits the number of accessible small modes along the line of sight -- thus making the forecasted sensitivity weaker. To better understand the extent to which these factors affect results for $\fNL$, it is instructive to see how well each of the five galaxy samples of SPHEREx constrain $\fNL$. Table \ref{SPHEREx_single_tracer_constraint} lists the single tracer results obtained from Fisher matrix calculations for each of the five galaxy samples of SPHEREx. From table \ref{SPHEREx_single_tracer_constraint}, we can see that higher density galaxy samples (samples 4 and 5) are the most sensitive when it comes to local $\fNL$. These forecasts are, however, also a bit degraded by the relatively large $\tilde{\sigma}_{z}$ of these samples -- but, in general, the higher density galaxy samples provide a stronger sensitivity to $\fNL$ in a galaxy power spectrum forecast.

\begin{table}[]
    \centering
    \begin{tabular}{||c|c|c||} 
    \hline
    $ \text{sample} $ & $\tilde{\sigma}_{z} $ & $\sigma(\fNL)$\\ [0.5ex] 
    \hline
    1 & 0.003 & 3.83 \\ [0.5ex]
    \hline
    2 & 0.01 & 2.34\\ [0.5ex]
    \hline
    3 & 0.03 & 2.05\\ [0.5ex]
    \hline
    4 & 0.1 & 1.48\\ [0.5ex]
    \hline 
    5 & 0.2 & 1.16\\ [0.5ex]
    \hline
    \end{tabular}
    \caption{SPHEREx galaxy samples and their respective single tracer results for $\fNL$ around fiducial $\fNL=1$.}
    \label{SPHEREx_single_tracer_constraint}
\end{table}

 \subsection{Effect of Neutrinos}
\begin{figure}[h!]
    \centering
    \includegraphics[width=\columnwidth]{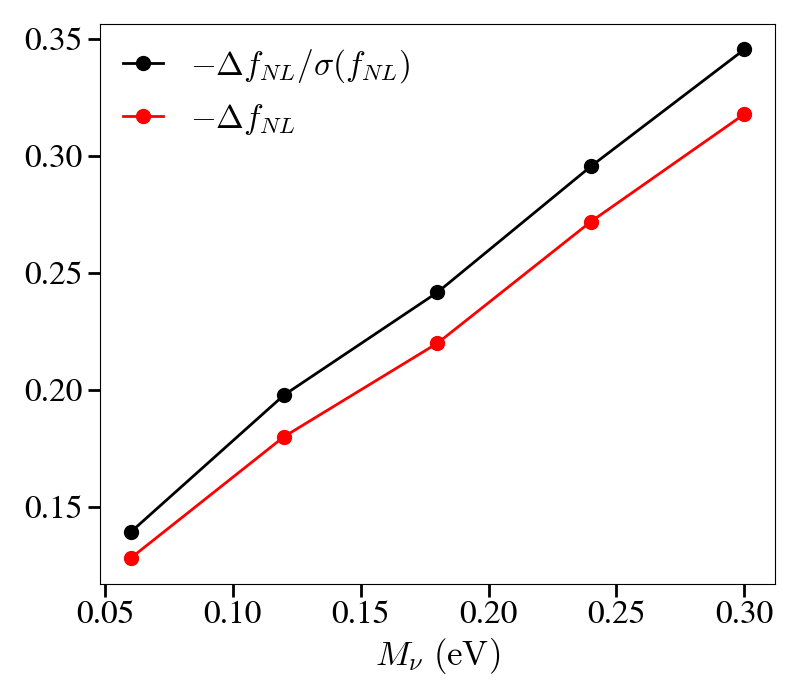}
    \caption{Absolute and relative bias of $\fNL^{\text{local}}$ upon ignoring the scale-dependent bias induced by neutrinos. Note that $\sigma(\fNL)\approx 0.9$ irrespective of the neutrino mass.  The biases plotted above are computed using our MCMC likelihood pipeline and use all five SPHEREx galaxy samples.}
    \label{fig:fNL_bias_neutrinos_all}
\end{figure}
\begin{figure}[h!]
    \centering
    \includegraphics[width=\columnwidth]{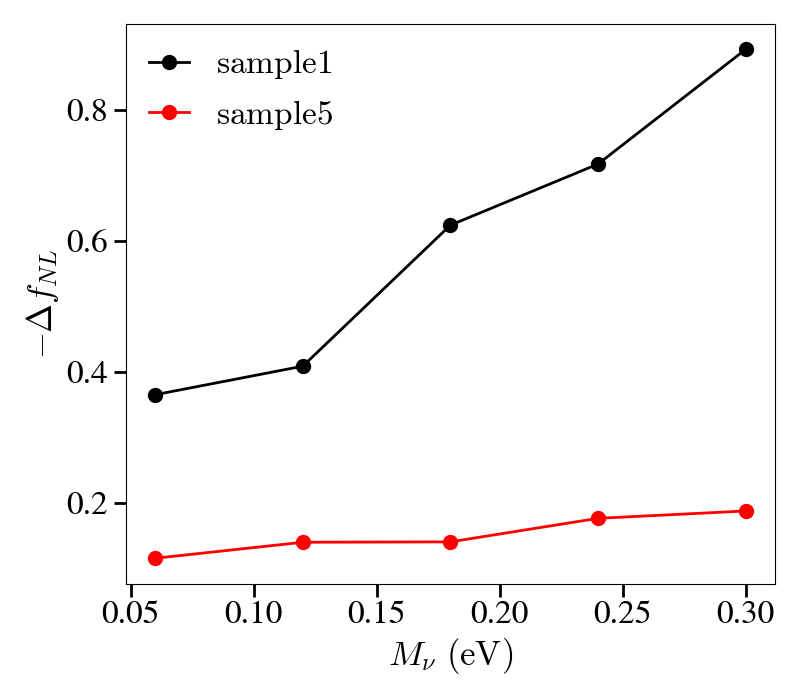}
    \caption{Absolute bias on $\fNL^{\text{local}}$ from ignoring neutrino scale-dependent bias for SPHEREx galaxy samples 1 and 5. }
    \label{fig:fNL_bias_neutrinos_15}
\end{figure}

\begin{figure}[h!]
    \centering
    \includegraphics[width=\columnwidth]{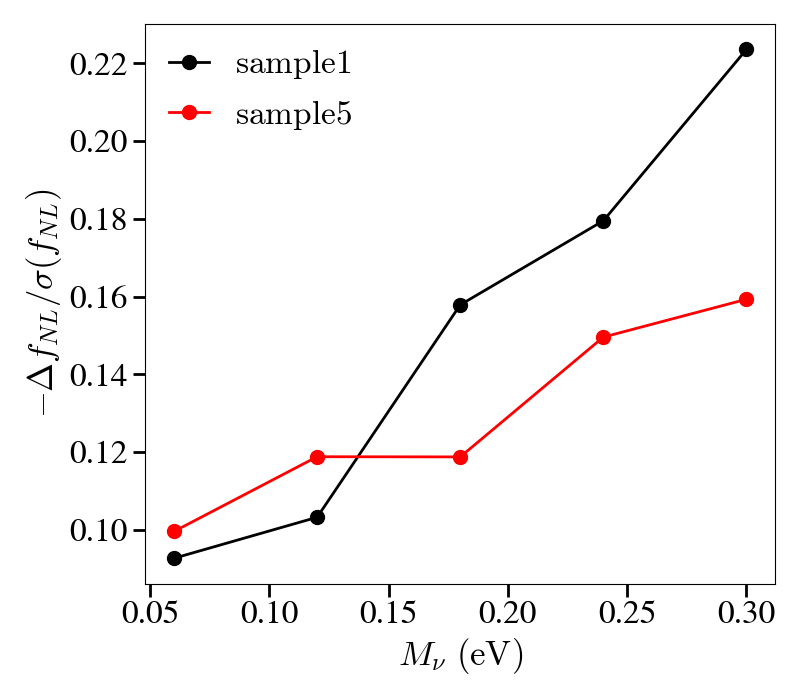}
    \caption{Relative bias in the MCMC forecast of $\fNL^{\text{local}}$ upon ignoring the scale-dependent bias induced by neutrinos of different masses for SPHEREx galaxy samples 1 and 5.  }
    \label{fig:fNL_rel_bias_neutrinos_15}
\end{figure}

Figure \ref{fig:dlogPg_neutrinos} shows that the free-streaming of neutrinos suppresses the galaxy bias on the largest scales. Ignoring this effect therefore leads to a negative bias in the measurement of $\fNL$. This is reflected in our MCMC forecasts where we observe that for neutrinos of total mass $M_{\nu}=0.06\ \text{eV}$, the forecast mean of $\fNL$ is biased by $\Delta \fNL=-0.14\sigma$. Figure \ref{fig:fNL_bias_neutrinos_all} shows the (negative) absolute ($-\Delta \fNL$) and relative ($-\Delta \fNL/\sigma(\fNL)$) biases in $\fNL$ for different total neutrino masses. The fact that the scale-dependence imparted by neutrinos increases with neutrino mass (as explained in Ref. \cite{Shiveshwarkar:2020jxr}) explains the increase in the (negative) absolute bias in $\fNL$ as seen in figure \ref{fig:fNL_bias_neutrinos_all}. In addition, we find that the effect of neutrinos does not significantly alter the forecast for $\fNL$ when properly included in the modelling of galaxy spectra. Irrespective of the neutrino mass, we obtain $\sigma(\fNL) \approx 0.9$ around $\fNL=1.0$ when fitting simultaneously for the effect of neutrinos, which is the same $\sigma(\fNL)$ we obtain in the absence of any large-scale effects due to neutrino/ionising radiation. This is why the relative bias $-\Delta(\fNL)/\sigma(\fNL)$ shows the same trend as the absolute bias in $\fNL$ in figure \ref{fig:fNL_bias_neutrinos_all}.

Figures \ref{fig:fNL_bias_neutrinos_15} and \ref{fig:fNL_rel_bias_neutrinos_15} show the (negative) absolute and relative biases in $\fNL$ obtained in single-tracer MCMC forecasts using our likelihood for the low density and high density samples (samples 1 and 5 respectively) of SPHEREx as function of neutrino mass. Figures \ref{fig:fNL_bias_neutrinos_all} and \ref{fig:fNL_rel_bias_neutrinos_15} show that for neutrinos with masses according to the minimal mass inverted hierarchy ($M_\nu \approx 0.1$eV), one can potentially obtain a (negative) bias between $0.1\sigma$ and $0.2\sigma$ when measuring $\fNL$ using galaxy power spectra. But at the same time, including the effect of neutrino-induced scale-dependent bias does not significantly worsen prospects for measuring $\fNL$. This makes the free-streaming of neutrinos an important systematic effect to take into account for future galaxy surveys that plan to measure $\fNL^{\text{local}}$ with increasing precision $(\sigma(\fNL)\lesssim 0.9)$. It is worth keeping in mind that if neutrino masses are larger, the bias from neutrinos is larger. Planck found a tight bound on the neutrino mass sum~\cite{Planck:2018vyg}, but some authors argue that the Planck lensing anomaly causes Planck constraints to be forced to low neutrino mass values, while reality may be different (see, e.g., Ref.~\cite{Sgier:2021bzf}, which found $M_{\nu}=0.51^{+0.21}_{-0.24}$; other analyses using $A_L$ find similarly larger values, see, e.g., Ref.~\cite{DiValentino:2021imh}).

 \subsection{Effect of Ionising Radiation}


The primary effect of ionising radiation fluctuations is a suppression of the galaxy power spectrum on scales larger than the mean free path of UV photons (Eq. \eqref{Pg_with_bJ}). 
However, the shot noise contribution in Eq. \ref{Pg_with_bJ} -- which is important if quasars have long lifetimes -- can temper this suppression, as it peaks at the largest scales and becomes more important at higher redshifts.  

\begin{table}[]
    \centering
    \begin{tabular}{||c|c|c|c|c||} 
    \hline    
    \multicolumn{5}{||c||}{ } \\ [-1.0ex]
    \multicolumn{5}{||c||}{$\sigma(\fNL)$} \\ 
    \multicolumn{5}{||c||}{ } \\ [-1.0ex]
    \hline
    $\text{SPHEREx}$ & $ $ & $P_{Jshot}=0$ & $t_{Q}=100\, \text{Myr}$ & $t_{Q}=\infty$ \\ 
    $\text{sample}$ & $b_{J}=0$ & $b_{J}=0.05$ & $b_{J}=0.05$ & $b_{J}=0.05$ \\[0.5ex]
    \hline
    1 & 3.83 & 6.47 & 6.42 & 5.77 \\ [0.5ex]
    \hline
    2 & 2.34 & 3.71 & 3.68 & 3.24 \\ [0.5ex]
    \hline
    3 & 2.05 & 3.14 & 3.12 & 2.70 \\ [0.5ex]
    \hline
    4 & 1.48 & 2.08 & 2.06 & 1.80 \\ [0.5ex]
    \hline
    5 & 1.16 & 1.44 & 1.42 & 1.26 \\ [0.5ex]
    \hline
    \end{tabular}
    \caption{Single tracer Fisher results for $\fNL$ in the presence of ionising radiation effects for different models of the quasar shot noise -- $t_{Q}$ in the above table is the quasar lifetime in the shot noise model. All the forecasts are around a fiducial $b_{J} = 0.05$, which is taken to be an independent parameter for all galaxy samples and redshifts.}
    \label{bJ_single_tracer_forecasts}
\end{table}

\begin{table}[]
    \centering
    \begin{tabular}{||c|c|c|c||} 
    \hline    
    \multicolumn{4}{||c||}{ } \\ [-1.0ex]
    \multicolumn{4}{||c||}{$\sigma(\fNL)$} \\ 
    \multicolumn{4}{||c||}{ } \\ [-1.0ex]
    \hline
    $\sigma_{\text{prior}}(b_{J})$ & $P_{Jshot}=0$ & $t_{Q}=100\, \text{Myr}$ & $t_{Q}=\infty$ \\[0.5ex] 
    $ b_{J}=0.05 $ & $b_{J}=0.05$ & $b_{J}=0.05$ & $b_{J}=0.05$ \\[0.5ex]
    \hline
    0.0 & 0.89 & 0.89 & 0.90  \\ [0.5ex]
    \hline
    0.025 & 0.90 & 0.90 & 0.91 \\ [0.5ex]
    \hline
    0.05 & 0.92 & 0.92 & 0.2  \\ [0.5ex]
    \hline
    0.1 & 0.96 & 0.95 & 0.94  \\ [0.5ex]
    \hline
    \end{tabular}
    \caption{Single tracer Fisher results for $\fNL$ in the presence of ionising radiation effects for different priors on the intensity bias and different models of the shot noise (see figure \ref{fig:sig_fNL_bJ_priors}) -- $t_{Q}$ in the above table is the quasar lifetime in the shot noise model. All the forecasts are around fiducial values of $b_{J} = 0.05$ for all galaxy samples and redshifts. There is no significant degradation of sensitivity upon marginalising over the intensity bias as long as we remember to impose reasonable priors on the $b_{J}$ reflecting the fact that $b_{J}\lesssim 0.1$.}
    \label{bJ_prior_forecasts}
\end{table}

Since the sensitivity to $\fNL$ can be affected by possibly different $b_J$ values, we now investigate the cost of marginalization over $b_J$, taking independent values of $b_J$ for each of our redshift bins and over all of our samples, since $b_J$ can be different in all of them. We note that this is maximally conservative. 
Marginalising over different $b_{J}$ at different redshifts without any priors in the single-tracer Fisher analysis was done in Ref. \cite{Sanderbeck:2018lwc} and results in a $40\%$ increase in $\sigma(\fNL)$. For individual SPHEREx galaxy samples we find a result that is qualitatively consistent with their findings:  We find that the extent of the degradation in the Fisher forecast of $\fNL$ upon marginalising over $b_{J}(z)$ varies with the tracer population. This is unsurprising as at fixed $b_J$ the tracer bias $b$ helps set the relative amplitude of the $\fNL$-dependent and UV background-dependent terms in the galaxy power spectrum. Table \ref{bJ_single_tracer_forecasts} shows the single tracer Fisher results for $\fNL$ for each SPHEREx galaxy sample obtained in the presence of UV effects and different quasar shot noise models. Table \ref{bJ_single_tracer_forecasts} shows that the degradation in the sensitivity to $\fNL$ is the highest for the low-density sample (sample 1, about 50\% effect) and lowest for the high-density sample (sample 5, 10-30\% depending on the lifetime model). Also, the sensitivity is less degraded (across all samples) in the presence of the maximal estimate of shot noise. We observe a similar trend in multitracer Fisher forecasts for $f_{\rm NL}$ -- the multitracer Fisher sensitivity obtained for samples 4 and 5 of SPHEREx is worsened by about $25\% $ when marginalising over different $b_J(z)$ for different galaxy samples. This multitracer sensitivity is less degraded in the presence of the maximal estimate of the shot noise (in which case the degradation is closer to $22\%$).
\begin{figure}[h!]
    \centering
    \includegraphics[width=\columnwidth]{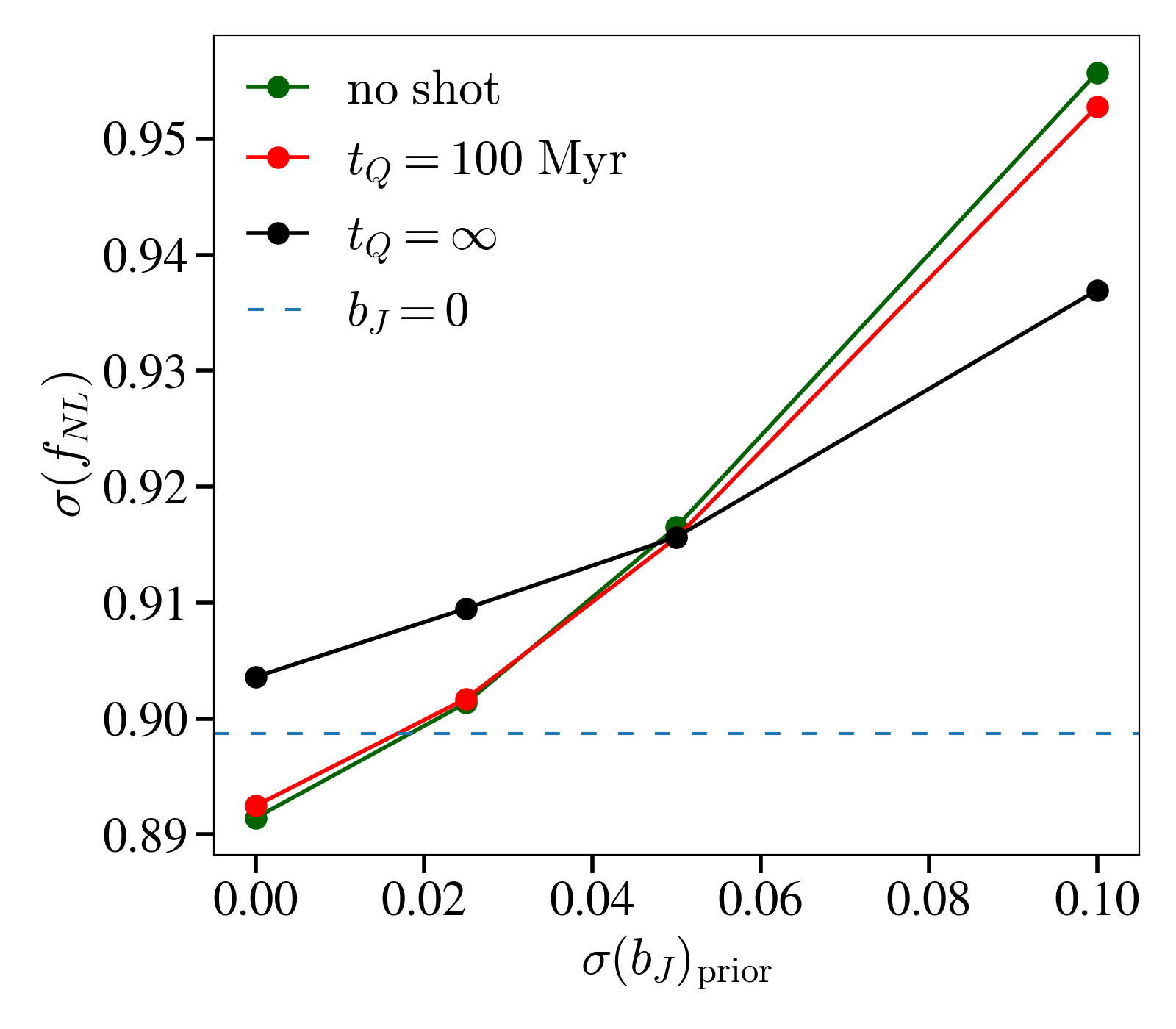}
    \caption{Multitracer Fisher results obtained in the presence of ionising radiation for all five SPHEREx galaxy samples for different priors on $b_{J}$ around fiducial values of $b_{J}=0.05$ for all galaxy samples and redshifts (see table \ref{bJ_prior_forecasts}). The dashed line shows the forecasted SPHEREx Fisher sensitivity on $\fNL$ in the absence of any non-primordial large-scale bias effects.} 
    \label{fig:sig_fNL_bJ_priors}
\end{figure}

These results suggest a significant cost to marginalising over the $b_J$ in the absence of any priors over the intensity bias.  Astrophysical considerations elaborated in Ref. \cite{Sanderbeck:2018lwc} suggest that the intensity bias $b_{J}$ at any redshift is smaller than $0.1$, and it could be much smaller. This makes it important to impose priors on the $b_{J}$ parameters when incorporating the effect of UV fluctuations. Upon doing so, we find that multitracer Fisher sensitivity to $\fNL$ obtained with all five galaxy samples of SPHEREx \textit{does not} differ significantly from the Fisher sensitivity obtained in the absence of any neutrino/ionising radiation effect (see table \ref{bJ_prior_forecasts}). Figure \ref{fig:sig_fNL_bJ_priors} and table \ref{bJ_prior_forecasts} show how $\sigma(\fNL)$ obtained from Fisher matrix analysis for all SPHEREx tracers in the presence of ionising radiation varies with the prior imposed on $b_{J}$. All the $b_{J}$ have a fiducial value of $b_{J}=0.05$. We see from figure \ref{fig:sig_fNL_bJ_priors} as well as from table \ref{bJ_prior_forecasts} that for reasonable priors on the $b_{J}$ i.e. $\sigma(b_{J})\lesssim 0.1$, the Fisher forecast of $\fNL$ is worsened by $\Delta\sigma \lesssim 5\%$. Thus, a reasonable prior on $b_{J}$, while including the effect of ionising radiation in modelling the galaxy power spectra, does not significantly weaken the galaxy power spectrum sensitivity forecast of local primordial non-Gaussianity with SPHEREx. 

\begin{figure}[h!]
    \centering
    \includegraphics[width=\columnwidth]{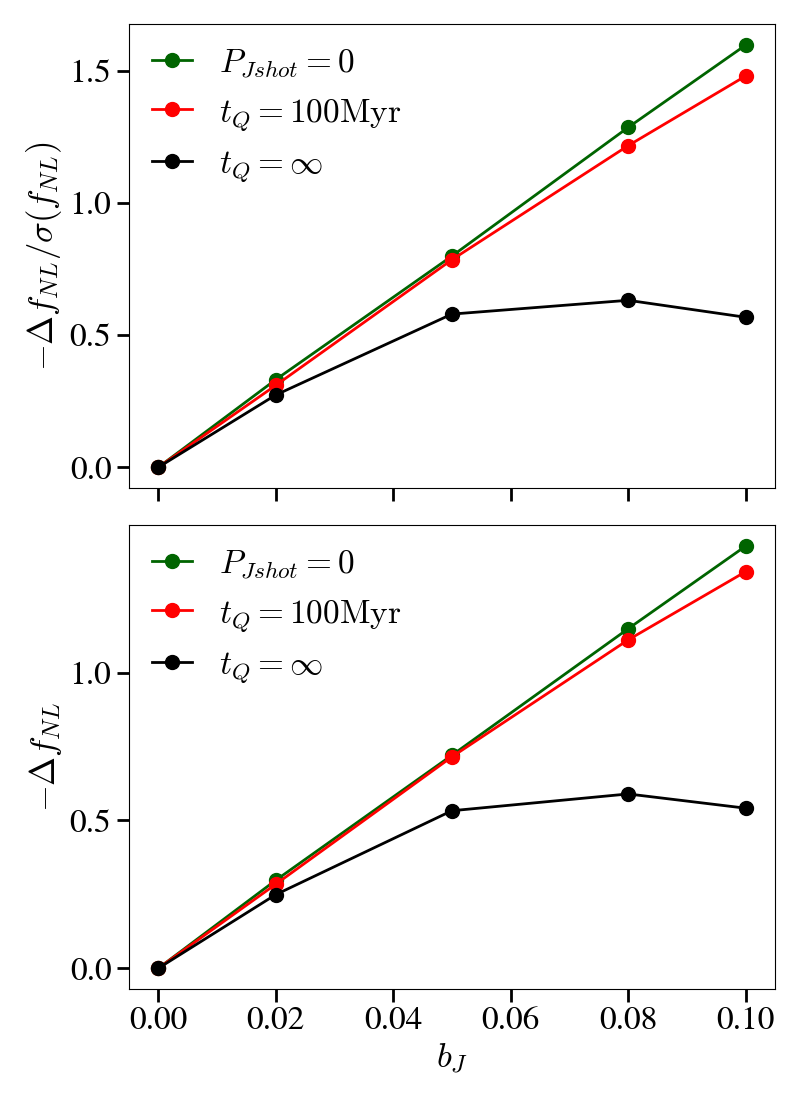}
    \caption{Absolute and relative bias in the MCMC forecast of $\fNL^{\text{local}}$ upon ignoring effect of ionising radiation for different intensity biases $b_{J}$ and shot noises corresponding to quasars of different lifetimes.}
    \label{fig:fNL_bias_bJ_all}
\end{figure}

On the other hand, ignoring the effect of ionising radiation while modelling the galaxy power spectra can significantly bias the measured value of $\fNL$. As mentioned in section \ref{ssec:UVbk}, a reasonable value for the intensity bias $b_{J}$ is between $0$ and $0.1$ but could vary significantly between samples. For simplicity of our forecasts for how $\fNL$ is biased by non-zero $b_{J}$, we assume that the intensity bias $b_{J}$ is the same for all galaxy samples of SPHEREx. As such, this is not a realistic model for the intensity bias, but gives a sense for the magnitude of bias that might be expected if intensity fluctuations are ignored.  Furthermore, since much of the constraining power on $\fNL$ comes from galaxy sample 5, we can roughly think of this as the bias that would be expected if sample 5 has intensity bias $b_{J}$. Note that in doing so, we obtain \textit{unmarginalised} forecast for $\fNL$ where we marginalise over all parameters except the intensity bias. The quasar shot noise $P_{Jshot}$ is in general scale-dependent (dependent on the mean free path of UV photons) and varies with the lifetime of quasars (see section \ref{ssec:UVbk} and figure \ref{fig:dlogPg_bJ_comparison}). We execute our MCMC analysis for quasars with infinite lifetime and a more realistic model with a lifetime of $100\, \text{Myr}$. We also do our analysis for the case of no shot noise, which is the limit of short lifetimes. 

Figure \ref{fig:fNL_bias_bJ_all} shows the absolute $(\Delta \fNL)$ and relative $(\Delta \fNL/\sigma(\fNL))$ bias in the MCMC forecast of $\fNL$ that we obtained for different estimates of the quasar shot noise for different values of the intensity bias $b_{J}$. The clustering of ionising radiation reduces power at larger scales and tends to negatively bias the forecast mean of $\fNL$. The general picture therefore is that ignoring the effect of ionising radiation on galaxy clustering leads to a negative bias in the measurement of $\fNL$ from the galaxy power spectrum. This negative bias is somewhat tempered at high intensity bias by the effect of quasar shot noise growing faster $(\propto b_{J}^{2})$ than the clustering term, which grows linearly with $b_{J}$ (see Eq. \eqref{Pg_with_bJ}).

Following figure \ref{fig:fNL_bias_bJ_all}, we see a clear increase in the (negative) absolute and relative bias with increasing $b_{J}$ for the models with no shot noise, corresponding to short quasar lifetimes, and with $100\, \text{Myr}$ quasar lifetimes. In these two models, the effect of the shot noise on the galaxy power spectrum remains subdominant even at high redshifts (see figure \ref{fig:dlogPg_bJ_comparison}) and the bias in $\fNL$ arises primarily because of the suppression in power at large scales causes by UV fluctuations. For the (unrealistic) infinite quasar lifetime model of the shot noise, however, the shot noise dominates at higher redshifts (figure \ref{fig:dlogPg_bJ_comparison}) and is large enough to offset the negative biasing effect of UV fluctuations at high $b_{J}$ -- this explains the turnaround in the bias around $b_{J}=0.08$ obtained with this model of shot noise. For $b_{J}=0.05$, it follows from \ref{fig:fNL_bias_bJ_all} that the (negative) bias in $\fNL$ can be anywhere between $0.5\sigma$ (maximal shot noise) and $0.8\sigma$ (no shot noise). Since quasar lifetimes likely are of the order of the Salpeter time of $10-100\, \text{Myr}$ \cite{Martini:2003ek}, the intensity bias can be as much as (if not bigger) than $1\sigma$ depending on the intensity bias. For our more realistic shot noise models, the bias in the measurement of $\fNL$ grows approximately linearly with the magnitude of the intensity bias $b_{J}$.

Fisher matrix calculations performed by Sanderbeck et al. in Ref. \cite{Sanderbeck:2018lwc} find that using an approximate fitting form for the UV transfer function $T_{J}$ going as $1/k$ (i.e. $T_{J}\propto 1/k$ or in other words, $P_{cJ}\propto P_{cc}/k$) removes much of the bias in $\fNL$ that results in ignoring ionising radiation fluctuations (without any quasar shot noise). This $1/k$ form was motivated by the UV transfer function on scales smaller than the mean free path of ionizing photons. These calculations were performed for redshifts up to $z=2.0$. MCMC analysis performed using our SPHEREx likelihood, which goes out to redshifts as large as $z\sim 4.5$, show that using a $1/k$ fitting form for $T_{J}$ actually leads to a \textit{positive} bias as large as $0.4\sigma$ in the forecast mean of $\fNL$ for $b_{J}=0.05$. This form does not work when the mean free path is shorter than the wavelength of modes that are being used, which is the case at high redshifts. It severely overestimates the UV transfer function at the largest scales at high redshifts $z\geq 2$. And since most of our $\fNL$ sensitivity comes from the largest scales, using a $1/k$ fitting form throughout the $k$-range overcompensates for the negative bias in $\fNL$ that results from ignoring the effect of UV fluctuations in the clustering of galaxies at high redshift ($z\geq 2.0$).  

Thus, to fit out for the effect of ionizing radiation requires a more sophisticated template than $P_{cJ}\propto P_{cc}/k$ for the cross-power spectrum of ionizing radiation with density.  Since the major ingredients that go into computing this cross-power spectrum are the mean free path of ionizing photons, which is precisely constrained ~\cite{Worseck:2014fya}, and the evolution of the sources of ionising radiation, which is well constrained by Ly$\alpha$ forest measurements~\cite{Haardt:2011xv}, we suspect that a next-order model over the simplistic $P_{cJ}\propto P_{cc}/k$ will be sufficient to remove most of the bias.  Fortunately, we found that if the template is sufficiently known, one can fit out the effect of ionizing radiation with little cost to the sensitivity to $f_{\rm NL}$.

\section{Discussion}
\label{sec:discussion}
With improved systematic control, observations of large-scale galaxy clustering provide a promising avenue to probe the physics of inflation. For squeezed limit primordial non-Gaussianity, such observations are especially important as the detection signal -- namely, the scale-dependent bias -- diverges at the largest scales.  Realistic galaxy surveys can only measure galaxy clustering up to distances about as large as the comoving Hubble radius today. Non-primordial physical effects due to free-streaming of light relics or ionising photons (generated at late times) can leave their imprints on the galaxy power spectrum at such \textit{near-horizon} scales, thereby contaminating the local primordial non-Gaussianity signal. Ignoring these effects when modelling galaxy power spectra can therefore bias the measurement of $\fNL$ -- thus rendering them important systematics for proposed surveys like SPHEREx and MegaMapper, which aim to constrain $\fNL$ to a precision of $\sigma(\fNL)\lesssim 1.0$. In this paper, we have investigated the impact of two such systematic effects -- the effect of free-streaming neutrinos and the effect of ionising radiation. Specifically, we performed MCMC forecasts for the measurement of local $\fNL$ around a fiducial value of $\fNL=1$ using the scale-dependent bias of galaxies obtained from the SPHEREx all-sky survey. We constructed a SPHEREx galaxy power spectrum likelihood within the MontePython parameter inference package~\cite{Audren:2012wb,MontePython3} and, together with a mock CMB power spectrum likelihood which \textit{does not} include any effects due to primordial non-Gaussianity\footnote{SPHEREx is expected to provide much more stringent constraints on $\fNL$ than CMB datasets -- consequently, we expect that including the effect of primordial non-Gaussianity in the CMB likelihood will not affect our results much.}, conducted MCMC runs with and without the aforementioned systematic effects incorporated in the galaxy power spectrum. We thus obtain a quantitative idea of the effect of said systematics on a precision measurement of $\fNL$ using the SPHEREx galaxy survey as a test case. 

The free-streaming of light relics/neutrinos induces a suppression of the galaxy bias on large scales following Eq. \eqref{bias_neutrinos_final} (see \ref{ssec:neutrinobk} and figure \ref{fig:dlogPg_neutrinos}). This effect (like other gravitational effects of free-streaming neutrinos) is sensitive to the \textit{sum} of neutrino masses and increases with the total neutrino mass~\cite{Shiveshwarkar:2020jxr}. As the scale-dependent bias induced by the free-streaming of neutrinos contaminates the $\fNL$ signal in the regime where it is most important (i.e. large scales), one might expect that the sensitivity to $\fNL$ would be somewhat degraded when it is taken into account. We find, however, that appropriately modelling the scale-dependent bias effect of neutrinos in the galaxy power spectrum leads to no significant degradation of $\sigma(\fNL)$ for three degenerate neutrinos with total masses up to $0.3\ \text{eV}$.

On the other hand, ignoring this effect can lead to a significant negative bias in the measurement of $\fNL$, especially for relatively more massive neutrinos. Figure \ref{fig:fNL_bias_neutrinos_all} shows the (negative) bias in the MCMC forecast of $\fNL$ due to three degenerate neutrinos of different masses (up to $M_{\nu} = 0.3\ \text{eV}$). Note that the extent of the bias in $f_{\rm NL}$ increases with increasing neutrino mass -- this is only to be expected because the scale-dependent effect of neutrino free-streaming on galaxy bias increases with the neutrino mass~\cite{Shiveshwarkar:2020jxr}. From figure \ref{fig:fNL_bias_neutrinos_all}, it is apparent that while this bias is about $0.1\sigma$ for neutrino masses equivalent to the minimal mass normal hierarchy, while it can be closer to $0.2\sigma$ for somewhat larger neutrino masses in the mass range also possible in an inverted hierarchy -- making this a relevant systematic for future measurements of $\fNL$, which aim for $\sigma(\fNL)\lesssim 1.0$. We also performed MCMC analyses for the SPHEREx high-density and low-density samples (samples 5 and 1, respectively) to investigate how the bias in $f_{\rm NL}$ could vary across different galaxy samples. Figures \ref{fig:fNL_bias_neutrinos_15} and \ref{fig:fNL_rel_bias_neutrinos_15} show that the absolute bias in $f_{\rm NL}$ resulting from ignoring the free-streaming of neutrinos is greater for the lower density sample. However, the lower density sample of SPHEREx also constrains $f_{\rm NL}$ quite weakly (see table \ref{bJ_single_tracer_forecasts}) -- so that the relative biases in $f_{\rm NL}$ are more or less of the same magnitude for both the high and low-density samples.

The effect of ionising radiation fluctuations (described in \ref{ssec:UVbk}) is the only astrophysical effect that is known to affect galaxy clustering on large scales~\cite{Sanderbeck:2018lwc}. The effect of ionising radiation fluctuations on the galaxy power spectrum (see Eq. \eqref{Pg_with_bJ}) is parametrized by the \textit{intensity} bias ($b_J$) per redshift bin and galaxy sample and the quasar shot noise ($P_{Jshot}$). The former encodes the response of the cooling rate of gas in dark matter haloes to the ambient fluctuations in ionising radiation energy density and leads to an effective suppression of galaxy bias at large scales. Calculations by Ref. \cite{Sanderbeck:2018lwc} show that any reasonable value of $b_J$ is between $0$ and $0.1$. On the other hand, the quasar shot noise (see section \ref{ssec:UVbk}) effectively increases the galaxy power spectrum at large scales and depends on the quasar lifetime ($t_{Q}$). Models with infinite quasar lifetime provide a maximal estimate for the shot noise~\cite{Sanderbeck:2018lwc} that can have the most impact on the forecast for $\fNL$, but even the $t_Q =100\, \text{Myr}$ model is likely extreme and so ignoring shot noise seems justified. Figure \ref{fig:dlogPg_bJ_comparison} shows the impact of both the matter-density tracing component of ionising radiation fluctuations and the quasar shot noise for different redshifts and different models of quasar lifetimes.

When incorporated in the modelling of the galaxy power spectra, there can be a significant cost to marginalising over the additional parameters in the galaxy power spectrum (i.e. the intensity biases) in terms of a degradation in the sensitivity to $\fNL$. Table \ref{bJ_single_tracer_forecasts} shows how single tracer Fisher sensitivity to $\fNL$ obtained for each galaxy sample of SPHEREx gets degraded by marginalising over the intensity biases (without any priors) for different estimates of the shot noise. These results are qualitatively consistent with the findings of Ref. \cite{Sanderbeck:2018lwc}, which show a $40\%$ degradation of single tracer Fisher sensitivity to $\fNL$ upon marginalising over the intensity biases. However, astrophysical considerations suggest the imposition of priors on the $b_{J}$ reflecting the fact that any realistic value of $b_J$ is thought to lie between $0$ and $0.1$. Imposing such reasonable priors on the $b_J$ changes the situation so that UV background fluctuations \textit{do not} significantly weaken the sensitivity to $\fNL$ obtained from the scale-dependent bias. Figure \ref{fig:sig_fNL_bJ_priors} shows how imposing reasonable priors on the $b_J$ does not significantly worsen the multitracer Fisher forecast of $\fNL$ obtained from galaxy power spectra for the SPHEREx survey. Thus, we must remember to impose reasonable priors on the intensity biases $b_{J}$ when incorporating the effect of ionising radiation in constraining local primordial non-Gaussianity from galaxy power spectra. 

On the other hand, not including the effect of ionising radiation in modelling the galaxy power spectra \textit{could} significantly bias the measurement of $\fNL$ from galaxy power spectra for $b_J\leq 0.1$. Figure \ref{fig:fNL_bias_bJ_all} shows the (negative) bias in the MCMC forecast of $\fNL$ for the SPHEREx survey for different models of the quasar shot noise and different values of the fiducial intensity bias $b_{J}$. For these MCMC runs, we fix the values of $b_J$ to be the same at all redshifts and for all galaxy samples of SPHEREx -- while this is not a realistic model, it gives an idea of how big a bias in $\fNL$ we would incur upon ignoring the effect of ionising radiation. Note that while the suppression of power caused by the UV transfer function ($T_{J}=P_{cJ}/P_{cc}$; see Eq. \ref{Pg_with_bJ}) in general leads to a negative bias in $\fNL$, the presence of quasar shot noise ($P_{Jshot}$) generically tempers this effect, especially for large quasar lifetimes and greater intensity biases. Except for the case of near-infinite quasar lifetimes, we note from figure \ref{fig:fNL_bias_bJ_all} that there is a monotonic increase in the (negative) bias in $\fNL$ with the intensity bias $b_J$. For the extreme upper limit on quasar lifetimes of $100\, \text{Myr}$, the negative bias in $\fNL$ can be as large as $1\sigma$ depending on the value of $b_J$. It is also likely to be greater for shorter quasar lifetimes as the quasar shot noise decreases with quasar lifetime. It therefore follows that an attempt to unambiguously detect local primordial non-Gaussianity by a precise measurement of $\fNL$ from galaxy clustering needs to incorporate the effect of ionising radiation while modelling galaxy power spectra. While modelling the effect of ionising radiation fluctuations, \citet{Sanderbeck:2018lwc} have shown that a fitting form for the UV transfer function going as $T_{J}\propto 1/k$ removes much of the bias in $f_{\rm NL}$ when constraining $f_{\rm NL}$ from the galaxy power spectra at redshifts $z\lesssim 2$. However, this fitting form overestimates the UV transfer function at the largest scales at higher redshifts and can lead to a \textit{positive} bias in the measurement of $f_{\rm NL}$. Consequently, we need a more sophisticated template than $T_{J}\propto 1/k$ to successfully fit for the UV transfer function at redshifts $z\gtrsim 2$.

In this paper, we have used the SPHEREx galaxy survey as a test case to quantify the importance of including some post-inflationary systematic effects. We contend that our results show the importance of including said effects in any measurement of local primordial non-Gaussianity from observations of large-scale galaxy bias. We conducted MCMC runs using our likelihood pipeline for the fiducial galaxy sample of the proposed MegaMapper survey as described in~\citet{Ferraro:2019uce}. We obtain an MCMC result of $\sigma(f_{\rm NL})=0.8$ from the galaxy power spectrum, which is consistent with the galaxy power spectrum forecast of \citet{Ferraro:2019uce}. Preliminary analysis for the MegaMapper galaxy sample using our likelihood shows that the negative bias in $f_{\rm NL}$ due to ignoring the scale-dependent bias induced by neutrinos can be as large as $0.25\sigma$ for total neutrino mass of $M_{\nu}=0.06\ \text{eV}$ (distributed among three degenerate species). Similarly, we find that ignoring the effect of ionising radiation can lead to a negative bias in $f_{\rm NL}$ as large as $2\sigma$ for the case of zero quasar shot noise and $b_{J}=0.05$. These biases in $f_{\rm NL}$ are significantly larger than the analogous biases for SPHEREx. While these are preliminary results, they still indicate that the systematic effects we investigate in this paper are likely to be more important for high-redshift surveys like MegaMapper~\cite{Ferraro:2019uce,Schlegel:2022vrv} that exclusively survey galaxies with redshift $z>2$. This is not surprising given that both the effect of neutrinos and ionising radiation generally increase with increasing redshift (see sections \ref{ssec:neutrinobk} and \ref{ssec:UVbk}). 

From our results, we also find that for the SPHEREx and MegaMapper galaxy samples, the effect of ionising radiation is larger than the effect of neutrino-induced scale-dependent bias in terms of how they bias a measurement of local primordial non-Gaussianity using the galaxy power spectrum. For instance, $|\Delta f_{\rm NL}|$ due to the effect of ionising radiation is as large as $0.8\sigma$ for the fiducial case of $b_{J}=0.05$, and scaling linearly with $b_J$ assuming zero quasar shot noise; whereas, even for the highest neutrino masses we investigate in our analysis, $|\Delta f_{\rm NL}|$ due to the free-streaming effect of neutrinos is no more than $0.4\sigma$. The difference is even starker for MegaMapper, where the typical (negative) bias in $f_{\rm NL}$ due to neutrinos is about $0.3\sigma$ while the (negative) bias due to ionising radiation can be more than $1\sigma$. Qualitatively, this just reflects the fact that the relative effect of ionising radiation on the galaxy power spectrum is larger than that of the free-streaming of neutrinos.

In principle, any attempt at constraining primordial physics from large-scale galaxy clustering observables needs to reckon (to varying degrees) with the kind of systematic effects we have investigated in this paper. In addition to the effects of free-streaming light relics and ionising radiation fluctuations we have analysed in this paper, general relativistic light cone effects can modulate the observed galaxy clustering on horizon scales and can masquerade as a primordial non-Gaussianity signal (see, e.g., Refs. \cite{Jeong:2011as,Koyama:2018ttg}). These lightcone effects can contaminate the primordial non-Gaussianity signal and should be modelled with a complete general relativistic treatment of galaxy bias~\cite{Yoo:2012se,Umeh:2019qyd,Umeh_2019}.

Moreover, there are additional parameters that can characterise non-Gaussianity in the primordial curvature perturbation and which can be detected using observations of galaxy power spectra. Notable among these are the $\gNL$ and $\tauNL$ type non-Gaussianities, which respectively generate a scale-dependent bias and a stochastic contribution to the galaxy power spectrum (see, e.g., Refs. \cite{LoVerde:2011iz,Smith:2011ub,Smith:2010gx}). Thus, future surveys such as SPHEREx and MegaMapper are also poised to constrain both $\gNL$ and $\tauNL$ type non-Gaussianities. Our galaxy power spectrum likelihood pipeline can be extended in a natural way to help obtain forecasts for such measurements, as well as to understand how they can be affected by post-inflationary systematics such as those investigated in this paper. Our galaxy power spectrum likelihood can also be naturally applied (with appropriate modifications) to investigate the viability of surveys such as SPHEREx and MegaMapper in observing imprints of large-scale features in the primordial power spectrum on galaxy clustering -- the detection of which can help constrain inflationary physics beyond the standard model of cosmology or particle physics~\cite{Slosar:2019gvt}. We leave all such investigations to future work.


\acknowledgements
We would like to thank Mikhail M. Ivanov and Chen Heinrich for useful discussions. We would like to thank Drew Jamieson for use of his publicly available CLASS code and Olivier Doré for providing information on the SPHEREx survey. TB thanks Martina Gerbino for helpful discussions. CS is grateful for support from the NASA ATP Award 80NSSC20K0541. ML is supported by the Department of Physics and the College of Arts and Sciences at the University of Washington and by the Department of Energy grant DE-SC0023183. TB was supported through the INFN project “GRANT73/Tec-Nu”, and by the COSMOS network (www.cosmosnet.it) through the ASI (Italian Space Agency)
Grants 2016-24-H.0 and 2016-24-H.1-201. This work is partially supported by ICSC – Centro Nazionale di Ricerca in High Performance Computing, Big Data and Quantum Computing, funded by European Union – NextGenerationEU. MM acknowledges support from NSF award AST-2007012. Results in this paper were obtained using the high-performance computing system at the Institute for Advanced Computational Science at Stony Brook University.

\appendix
\section{Explicit Expression for galaxy power spectra}
\label{app:fullspectra}
In this appendix, we give the explicit expression for both the auto- and cross-power spectra of galaxies in a multitracer survey such as SPHEREx, including the effect of ionising radiation fluctuations.

We begin with Eq. \eqref{eq:dg_with_bJ} for a particular galaxy sample: 
\begin{eqnarray}
    \delta_{g,i} = b_{g,i}\delta_{c}-b_{J,i}\delta_{J} \ ,
\end{eqnarray}
where $b_{g,i}$ are the \textit{net} galaxy biases for different galaxy samples. The galaxy biases $b_{g,i}$ include any effects of free-streaming light relics or, in general, any contributions to the matter bispectrum that survive in the nearly squeezed limit. On the other hand, $b_{J,i}$ are the intensity biases that encode the response of galaxy formation to ambient ionising radiation fluctuations. Squaring the above equation we get 
\begin{eqnarray}
P_{g,ij} = b_{g,i}b_{g,j}P_{cc}\left(1-\frac{b_{J,i}}{b_{g,i}}T_{J}\right)\left(1-\frac{b_{J,j}}{b_{g,j}}T_{J}\right) \\ \nonumber
+ b_{J,i}b_{J,i}P_{Jshot} \ ,
\end{eqnarray}
where $T_{J}=P_{cJ}/P_{cc}$ is the UV transfer function and $P_{Jshot}$ is the quasar shot noise (see subsection \ref{ssec:UVbk}). Accounting for redshift-space distortions, the Alcock Paczynski effect and other terms in the galaxy power spectrum model for the SPHEREx survey (see section \ref{sec:Pk}), we have the following expression for the theoretical galaxy power spectra, which enter the covariance matrix defined in Eq. \eqref{eq:data_cov_expression}:
\begin{eqnarray}
    P_{i,j}^{th} = f_{AP}f_{BF} f_{\sigma_{z}}
    P_{cc} \left(b_{g,i}(1+\beta\mu^2) - b_{J,i}T_{J}\right)\times   \\ \nonumber 
    \left(b_{g,j}(1+\beta\mu^2) -b_{J,j}T_{J}\right) + f_{AP}f_{BF}f_{\sigma_{z}}b_{J,i}b_{J,j}P_{Jshot} \ .
\end{eqnarray}

\newpage
\bibliographystyle{apsrev4-1}
\bibliography{references.bib}

\end{document}